\documentclass[journal]{IEEEtran}

\ifCLASSINFOpdf
  
\else
 
\fi

 \usepackage{amsmath} 
 \usepackage{algorithm}
 \usepackage{algpseudocode}
 \usepackage{graphicx}
 \usepackage{caption}
 \usepackage{subfigure}
 \usepackage{float}
 \usepackage{cite}
 \usepackage{hyperref}
 \usepackage{xcolor}
\usepackage[switch]{lineno}

\begin{document}
\title{Adaptive Laser Beam Engineering with Coherent Beam Combining for Efficient Power Delivery} 
\author{Khushboo Soni, 
S.~Thirumugam, John Rozario Jegaraj, Nithyanandan Kanagaraj
\thanks{Khushboo Soni, S.~Thirumugam, and Nithyanandan Kanagaraj 
(e-mail: nithyan@phy.iith.ac.in) are with the Department of Physics, 
Indian Institute of Technology, Hyderabad, Telangana, 502285, India.} 
\thanks{John Rozario Jegaraj (e-mail: johnmfrg.drdl@gov.in) is with the Defence Research \& Development Laboratory,  Kanchanbagh, Hyderabad-500058, Telangana, India.}}

\maketitle

\begin{abstract}
High-power laser technologies are essential in precision manufacturing, defense, and scientific research, where accurate control of the beam profile is paramount. Although several beam-shaping methods exist, they often face implementation and scalability challenges. To address these limitations, we introduce a comprehensive and versatile framework for on-demand beam engineering through coherent beam combining (CBC) systems to precisely craft far-field intensity distributions. The proposed approach integrates limitless key capabilities: (i) dynamic beam shaping through sequential steering, (ii) structured static beam shaping allowing the direct formation of target-defined profiles, and (iii) high-speed dynamic beam sequencing without mechanical movement. Thus, the proposed approach could be a potential one-stop solution to meet wide manufacturing requirements.  Rapid reconfiguration is achieved through optimized phase control of the CBC channels, supported by a deep-learning–inspired optimization algorithm. This unified CBC framework significantly improves beam uniformity, power delivery efficiency, and scalability compared to conventional techniques, thus establishing a robust platform for next-generation laser systems in industrial manufacturing, materials processing, and directed-energy systems.
\end{abstract}

\IEEEpeerreviewmaketitle

\section{\textbf{Introduction}}

The unique combination of precision, power, and controllability offered by fiber lasers has established them as foundational technologies with transformative impact across precision manufacturing, defense systems, biomedical photonics, energy transfer, and advanced optical research~\cite{shi2014fiber,svelto2009properties,shiner2005fiber}. This growing reliance creates a dual challenge: the need to scale output power alongside the need for advanced control over beam structure. To address this, advanced beam shaping has emerged as a key enabler of progress~\cite{dickey2018laser}. By tailoring the spatial and temporal characteristics of laser beams, beam shaping supports a wide range of applications~\cite{kim2024review,leger1997laser}—from micromachining and additive manufacturing~\cite{schmidt2017laser} to welding~\cite{brinkmeier2025influence}, drilling, surface functionalization, and directed energy systems—optimizing the interaction between the laser and the target material~\cite{bakhtari2024review}. As high-power laser systems continue to drive progress in modern manufacturing, precision engineering, and emerging photonic technologies, the demand for versatile and reconfigurable beam shaping methods has accelerated, positioning it as a critical capability for the next generation of laser-based applications \cite{galbusera2024analytical,zervas2014high}.
Merely scaling laser power does not ensure better performance; uncontrolled energy distribution can result in material damage, inefficiencies, and reduced process fidelity. In contrast, beam shaping provides structured and adaptive control of the laser output, enabling uniform energy deposition, reduced thermal effects, and more reliable interactions with materials. These advances directly enhance the efficiency, accuracy, and reproducibility of laser–material processes, highlighting beam shaping as a key driver for future progress in high-power laser technologies.


Historically, a wide range of strategies have been explored to manipulate laser beams for diverse applications. Early solutions relied on refractive optics~\cite{laskin2012imaging}, such as optical arrays~\cite{streibl1989beam}, aspheric lenses~\cite{hoffnagle2003beam}, or phase plates~\cite{de2017beam}, which could redistribute the beam into simple predefined patterns with high efficiency. These devices were valued for their simplicity and low losses, but they were inherently static: once fabricated, they could only deliver a fixed intensity profile, and modifying the output required replacing the expensive optical component.

To expand the design flexibility, diffractive optical elements (DOEs) were later introduced~\cite{katz2018using,khonina2025advancements}. By encoding microstructured phase profiles, DOEs enabled the generation of far more intricate beam patterns and opened the door to compact, application-specific designs~\cite{kayahara2025effects}. However, they shared key drawbacks with refractive elements: the lack of dynamic reconfigurability, high sensitivity to wavelength variations, and limited resilience under high optical power, which restricted their use in demanding industrial or defense contexts.

In parallel, mirror-based beam shaping systems were developed, including segmented mirrors and deformable mirrors~\cite{bremer2024design,feng2025design}. These offered a degree of tunability through active surface deformation or relative tilt control, enabling real-time adjustments of the wavefront. While effective for correcting aberrations and modest reshaping tasks, they were restricted by mechanical complexity, limited spatial resolution, and scalability challenges when more intricate beam synthesis was required.

A significant step forward came with the advent of spatial light modulators (SLM), based on liquid-crystal or micro-mirror arrays\cite{metel2018power,rosales2024structured}. SLMs introduced fully programmable phase and amplitude modulation, granting unprecedented versatility in generating arbitrary two-dimensional beam patterns. SLMs allow dynamic control of the wavefront and can create many different beam patterns. However, their use in high-power laser systems is limited by low/moderate damage thresholds, modest efficiency, small aperture sizes, relatively slow response times, and high cost.

\begin{figure*}[t]
    \centering
    \subfigure[]   
    {\includegraphics[width=0.27\linewidth]{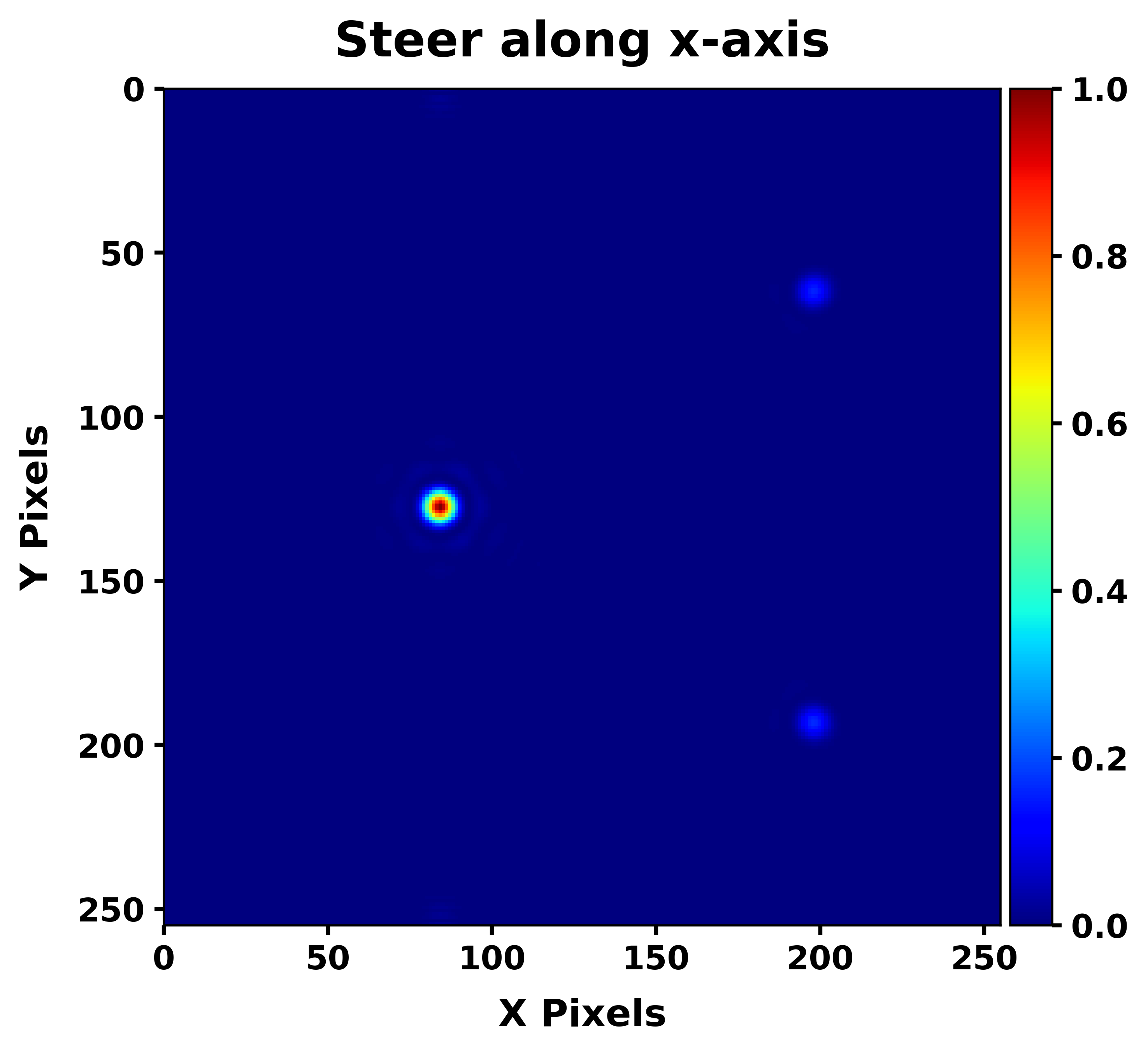}}
    \subfigure[]{\includegraphics[width=0.27\linewidth]{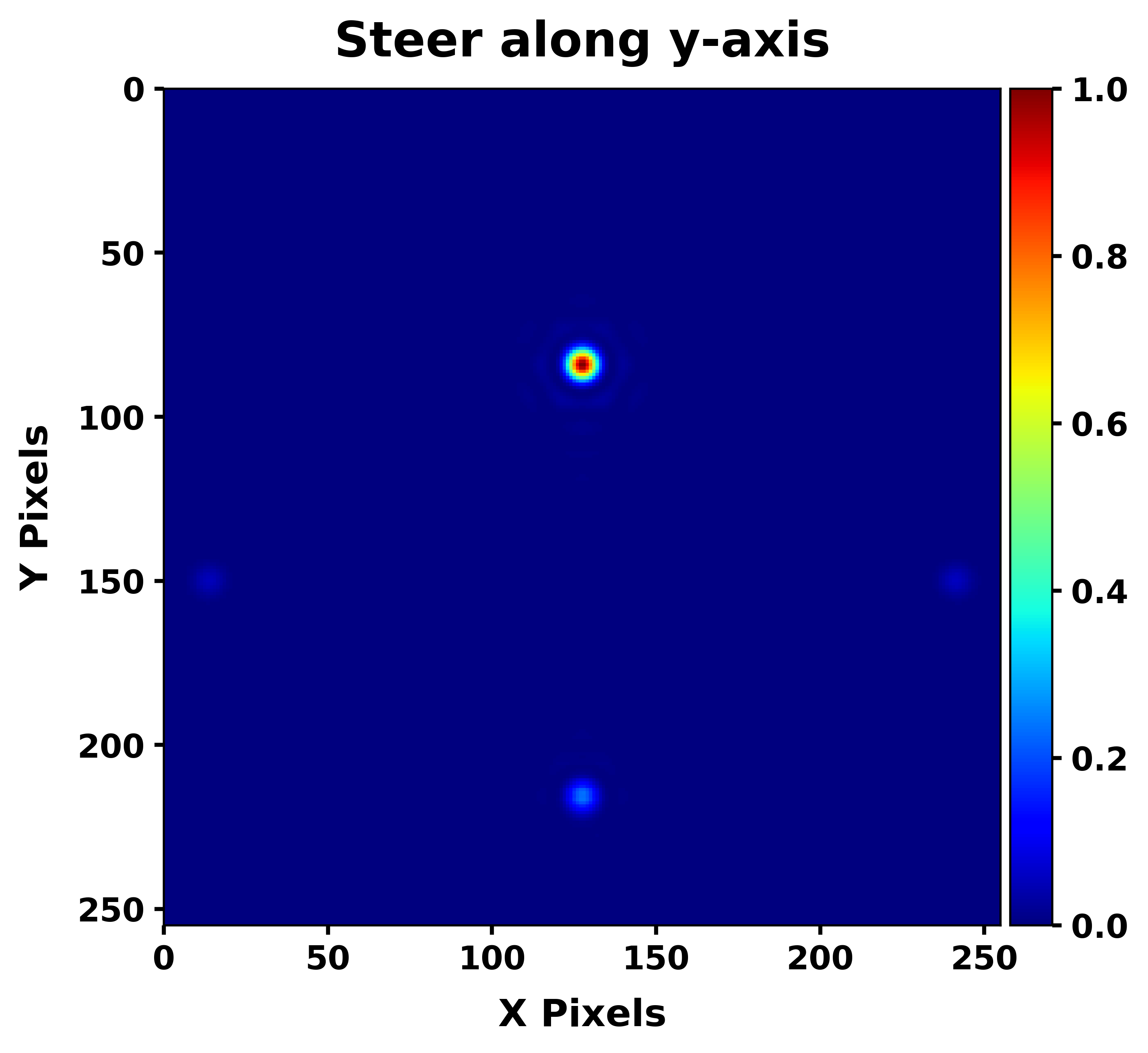}}
    \subfigure[]{\includegraphics[width=0.27\linewidth]{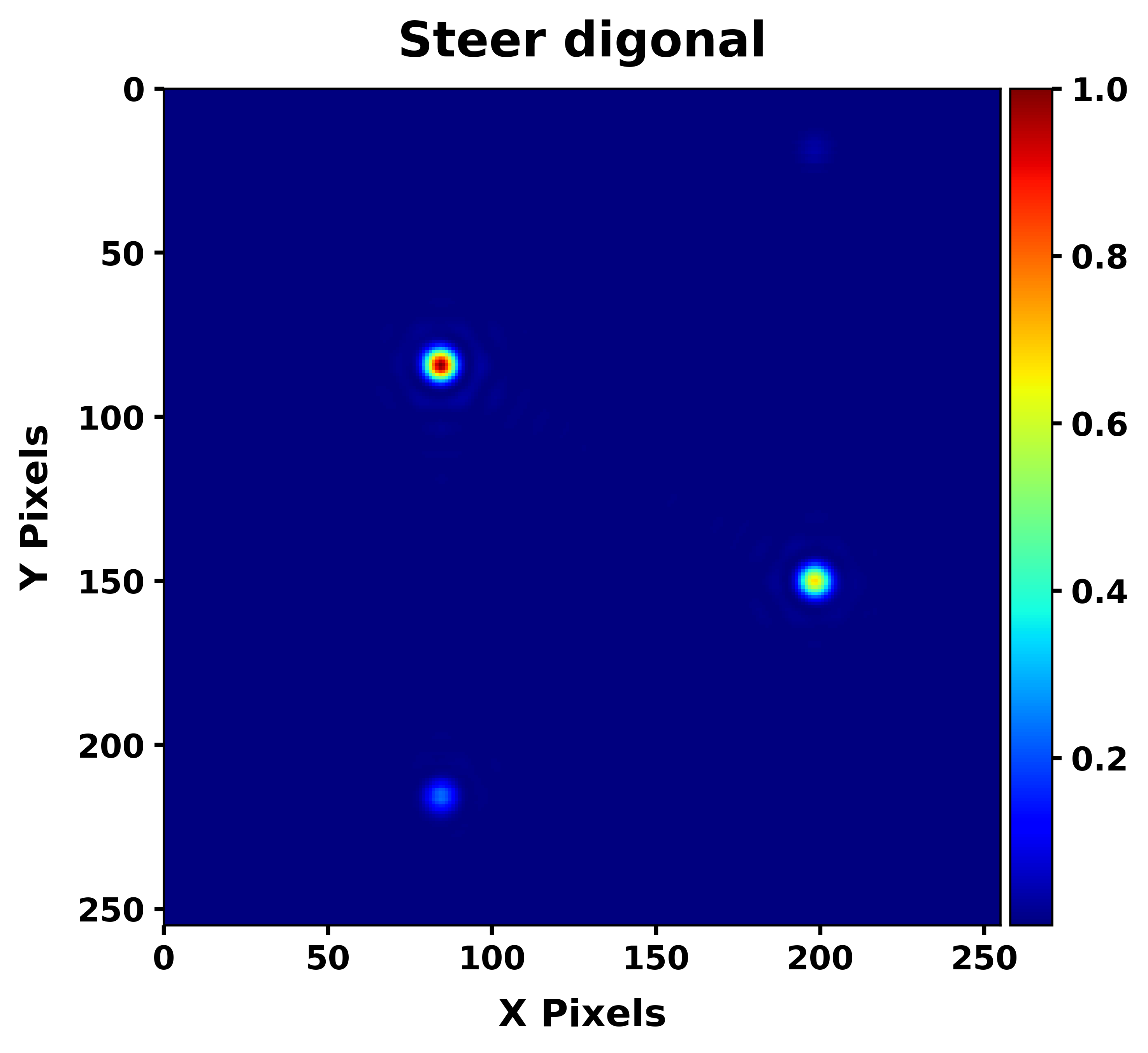}}

    \par\medskip 

    \subfigure[]{\includegraphics[width=0.27\linewidth]{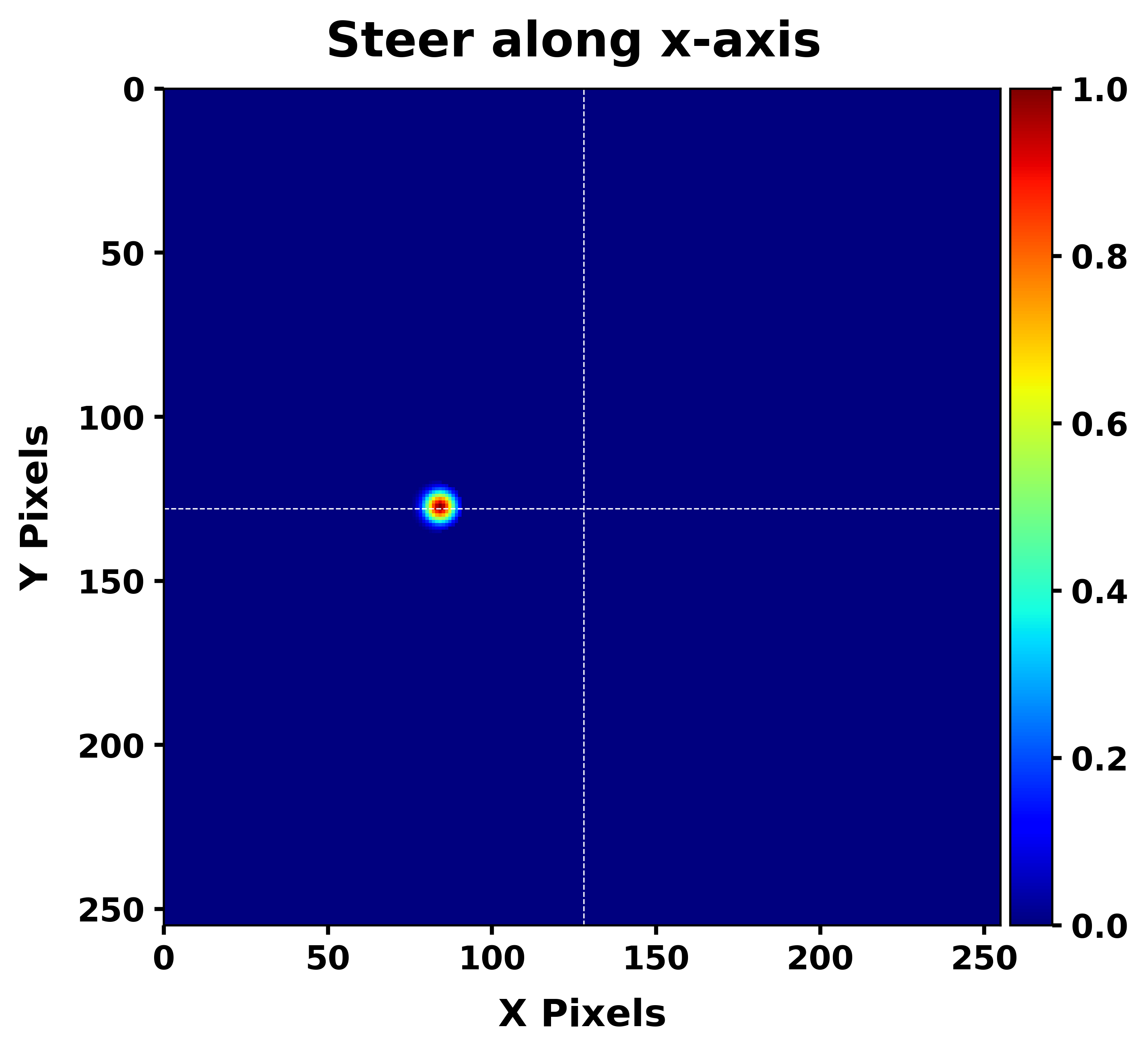}}
    \subfigure[]{\includegraphics[width=0.27\linewidth]{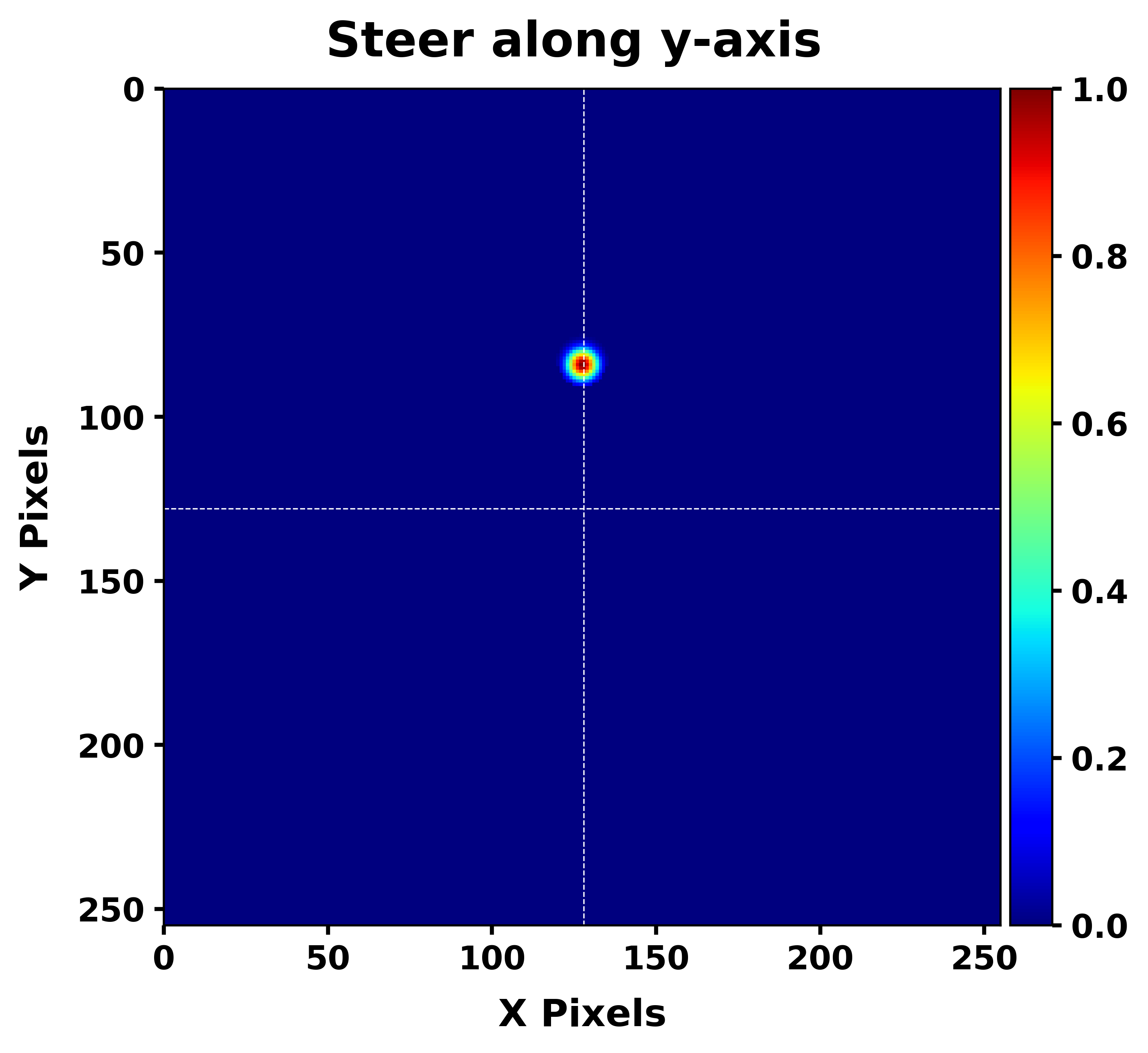}}
    \subfigure[]{\includegraphics[width=0.27\linewidth]{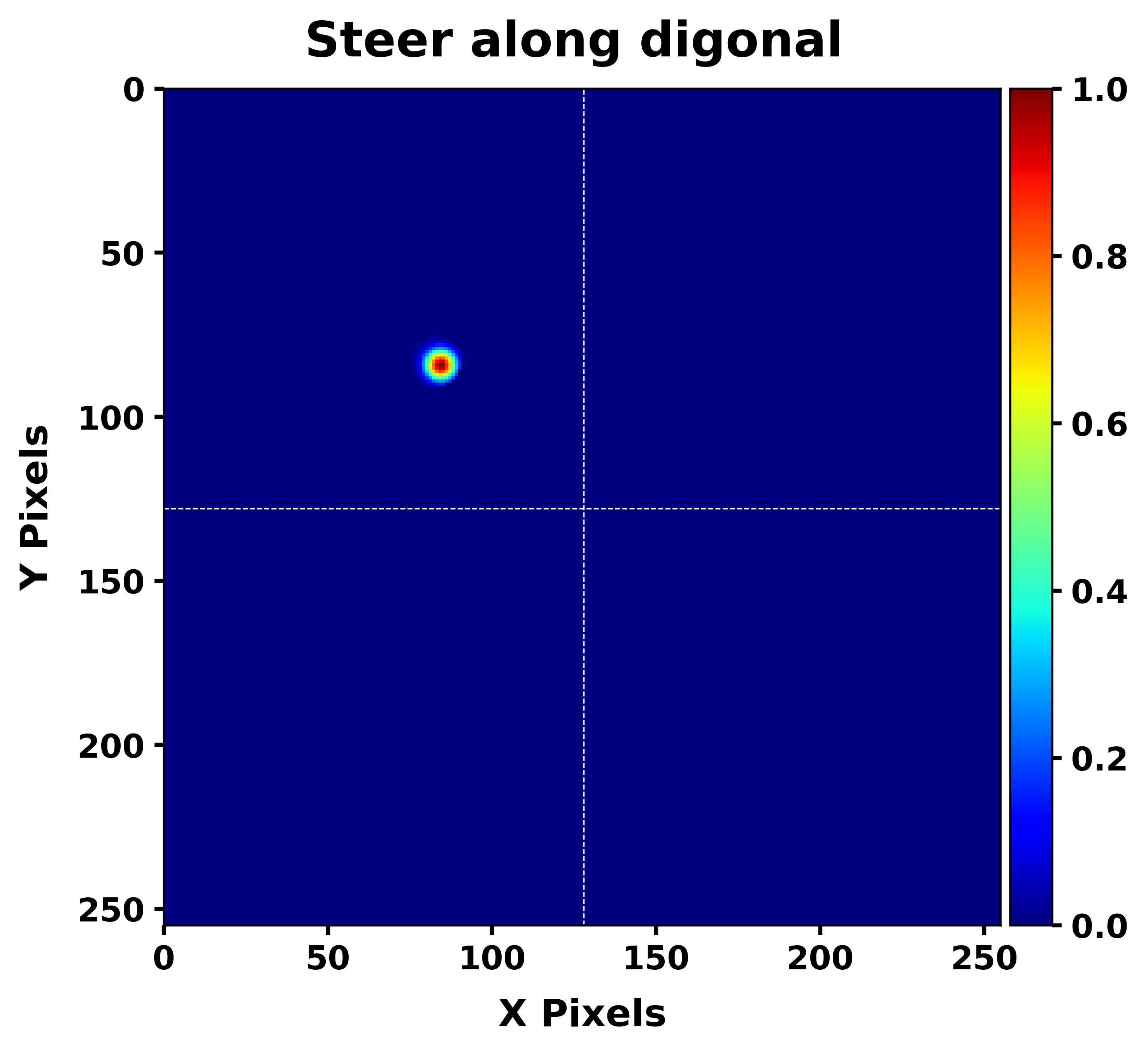}}

   \caption{\centering Beam steering at different locations: (a--c) with side lobes and (d--f) without side lobes. 
(a) \& (d) correspond to steering along the X-axis, 
(b) \& (e) along the Y-axis, 
and (c) \& (f) along the diagonal.}

    \label{Beam_steer_at_different_location}
\end{figure*}
More recently, metasurfaces and freeform optical elements have shown the ability to control light with very high precision~\cite{nikolov2021metaform}. By using micro- or nano-structured surfaces, they can bend, focus, or redistribute beams in compact and customized ways. Yet, challenges such as fabrication complexity, narrow operating bandwidth, and poor thermal stability under high power still limit their practical use in industrial settings\cite{gigli2021fundamental,feng2013beam}.

In summary, while many methods exist for beam shaping, most approaches lack to combine the key requirements of scalability, reconfigurability, efficiency, robustness, and high-power compatibility. These limitations highlight the need for new solutions that go beyond single-aperture modulation. 

One such promising approach is coherent beam combining (CBC), which coherently adds the outputs of multiple high-power fiber lasers through active phase control~\cite{brignon2013coherent,maji2025investigation,fathi2021towards,chang2020first,linslal2022challenges}. 
Instead of relying on a single aperture, CBC splits the output of a high-power fiber laser, amplifies it across multiple channels, and then actively phase-controls each sub-aperture before coherently combining the beams in the far field.
While CBC has traditionally been developed for power scaling, recent studies have also explored its potential for beam shaping and pattern control.
Adamov et al.~\cite{adamov2021laser} demonstrated amplitude–phase control in a fiber-laser array for static beam shaping; however, the method provided general field shaping but lacked flexibility for generating user-defined or highly structured target patterns.
Su et al.~\cite{su2024dynamic} extended this concept using a compact CBC configuration, but the resulting intensity distributions showed limited spatial uniformity and adaptability, especially for complex patterns. Recent advances, such as the Civan Laser DBL platform
Shekel~\cite{shekel2024dynamic} and Weber et al.~\cite{weber2025basic} have demonstrated high-speed dynamic beam shaping through coherent beam combining, where rapid phase tilting enables point-to-point steering and pattern generation at MHz rates. The approach primarily rely on sequential point switching rather than true phase-optimized shaping of continuous two-dimensional profiles. 
More recently, Chernikov et al.~\cite{chernikov2025deep} and Jabczyński~\cite{jabczynski2025analysis,mills2022single} investigated machine-learning-assisted phase retrieval and coherently generated field structures, offering theoretical insights but without addressing adaptive or user-defined beam shaping.
Consequently, static beam shaping, dynamic shape transitions, and quantitative analyses of power uniformity and spatial fidelity remain largely unexplored.

In this work, we present a self-contained and comprehensive framework for adaptive laser beam engineering using coherent beam combining. The proposed method unifies sequential steering, static beam shaping, and dynamic shape switching within a single phase-controlled architecture. By leveraging adaptive optimization, our approach enables precise formation and temporal evolution of user-defined beam patterns such as rings, rectangles, and triangles with uniform power distribution and rapid reconfigurability.

 This architecture offers a dual advantage: it not only scales the output power beyond the limitations of a single emitter but also enables programmable control over the far-field distribution by tuning the relative phases among array elements. In this way, CBC can electronically synthesize steerable beams, static beam shapes, and dynamic shape switching, all without the need for delicate high-power optical components. This ability to reconfigure profiles on demand while scaling power makes CBC particularly promising for advanced manufacturing, directed-energy applications, and next-generation photonic systems~\cite{chen2022partially}.


Despite these advantages, achieving robust and adaptive beam shaping through CBC requires precise, scalable, and efficient phase control. Conventional stochastic optimization methods, such as stochastic parallel gradient descent (SPGD) have proven effective for small arrays but often suffer from slow convergence and limited scalability~\cite{linslal2022challenges}. More recently, hybrid approaches and optimization algorithms inspired by machine learning have begun to show promise in overcoming these bottlenecks, offering faster adaptation and more accurate phase control across larger arrays~\cite{wang2021stabilization,mills2022single}.

Within this framework, we develop practical phase-optimization methods in CBC for high-power beam shaping, all implemented within the same system. The approach includes complementary capabilities: (i) dynamic beam shaping through sequential steering, which traces complex patterns; (ii) static beam shaping, which directly forms target profiles such as rings, rectangles, and triangles; and (iii) high-speed beam sequencing, which rapidly switches among precomputed shapes without mechanical motion. As all three approaches operate on the same CBC platform, the beam-shaping method that is suitable can be selected based on the specific manufacturing or processing requirement. Together, these capabilities offer a flexible and efficient solution for high-power beam control in advanced manufacturing, directed-energy systems, and modern laser applications.
\begin{figure*}[t]
    \centering
    \subfigure[]{\includegraphics[width=0.24\linewidth]{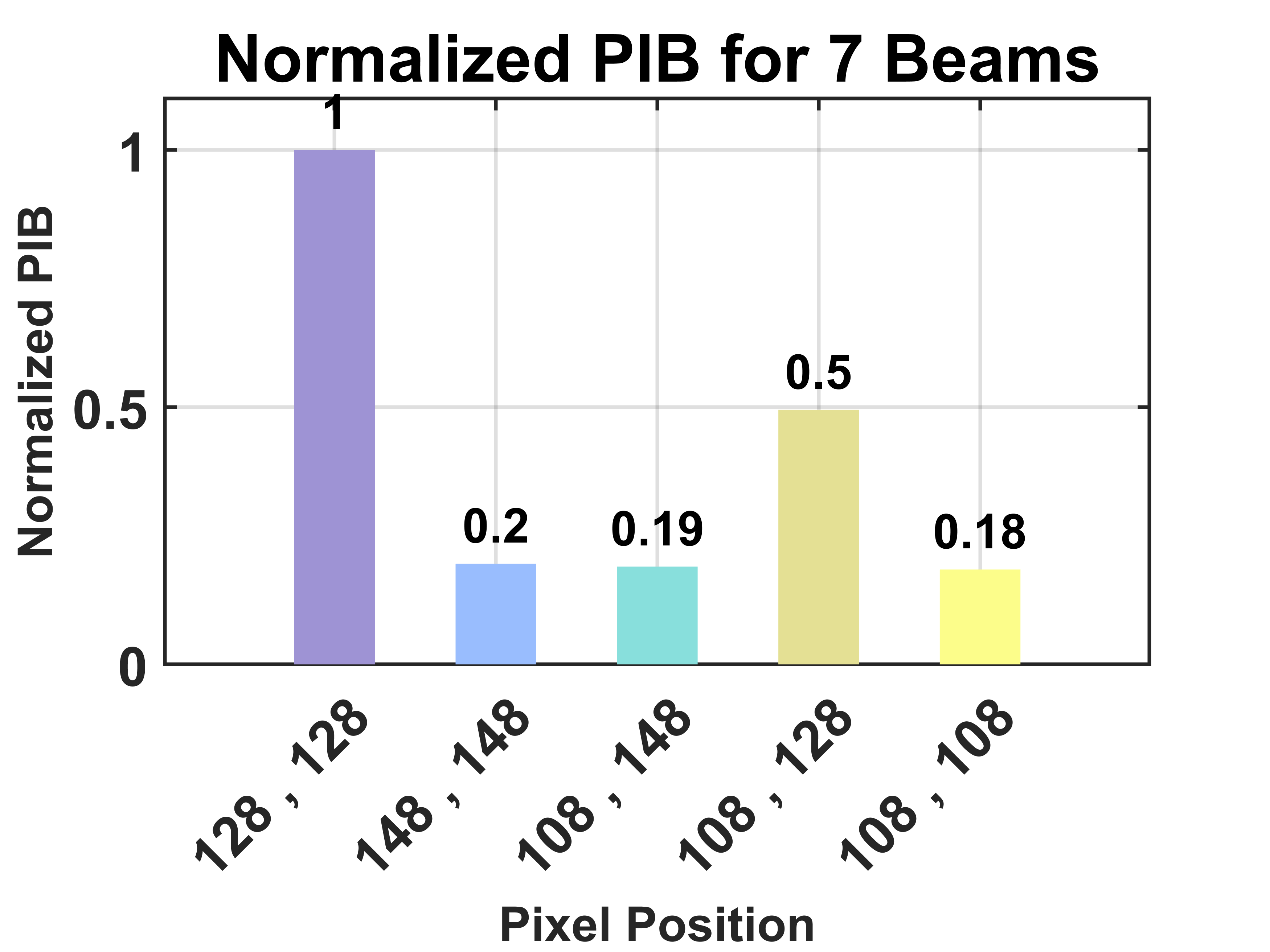}\label{fig:PIB_7}}
    \subfigure[]{\includegraphics[width=0.24\linewidth]{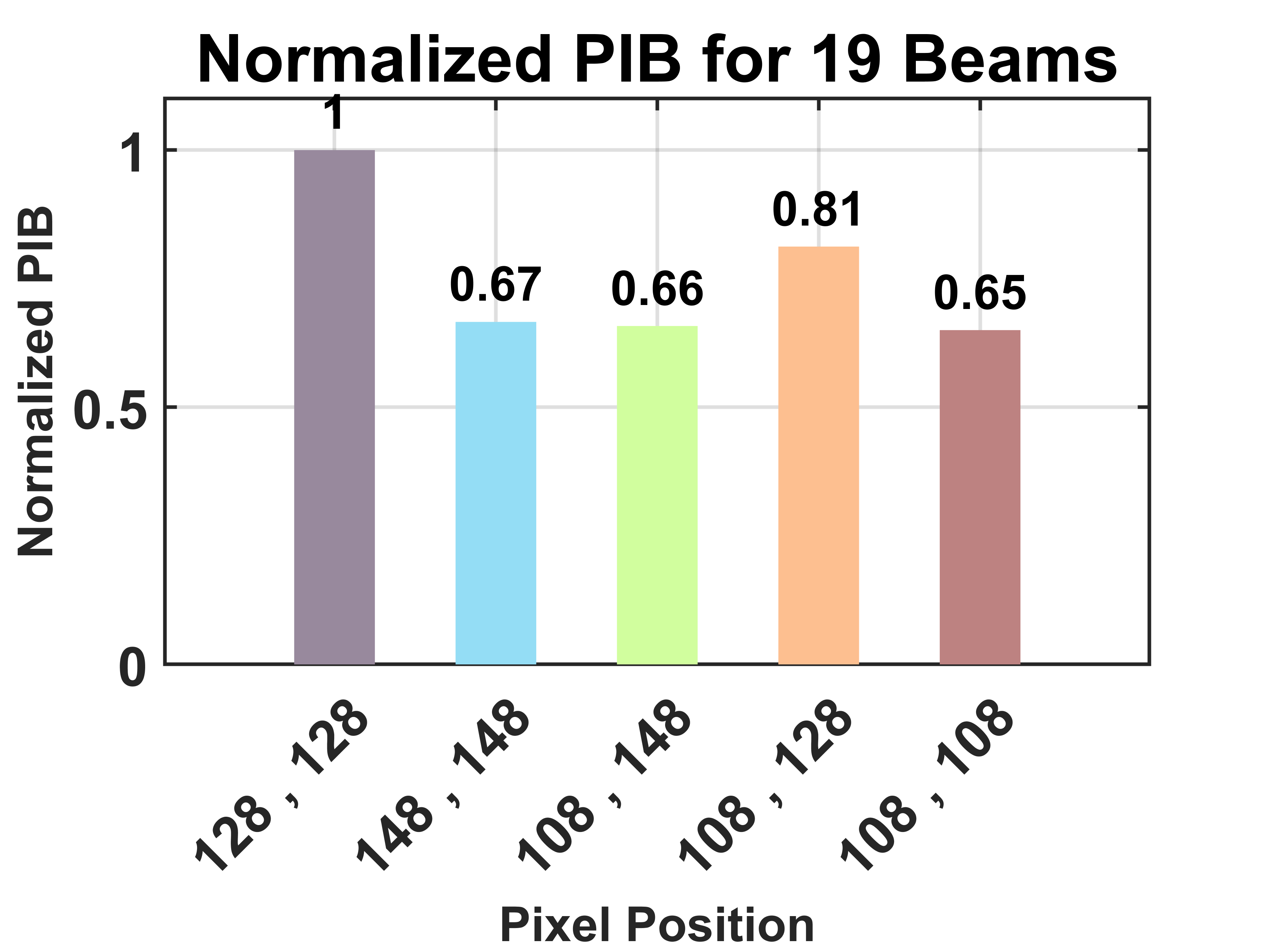}\label{fig:PIB_19}}
    \subfigure[]{\includegraphics[width=0.24\linewidth]{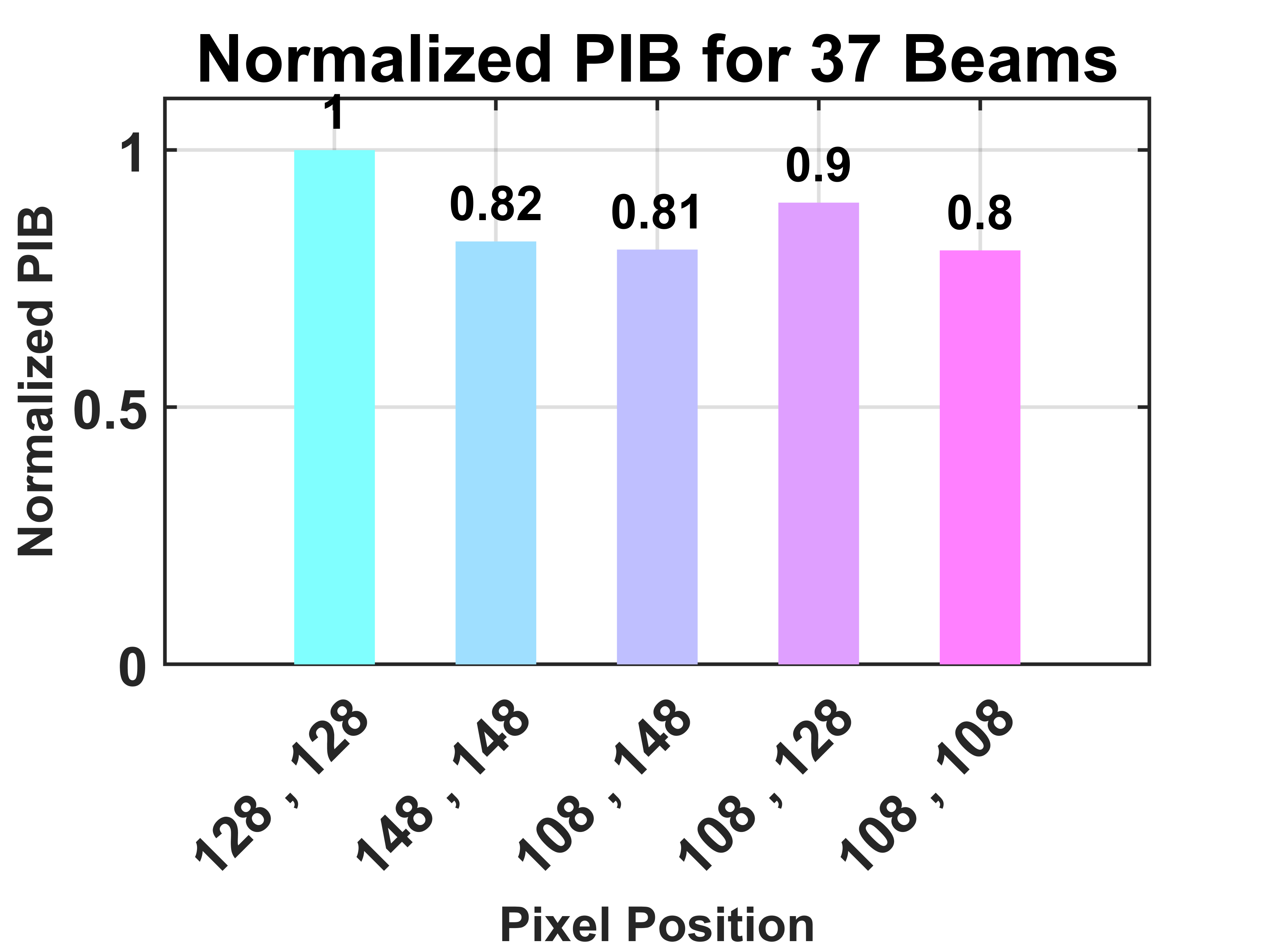}\label{fig:PIB_37}}
    \subfigure[]{\includegraphics[width=0.24\linewidth]{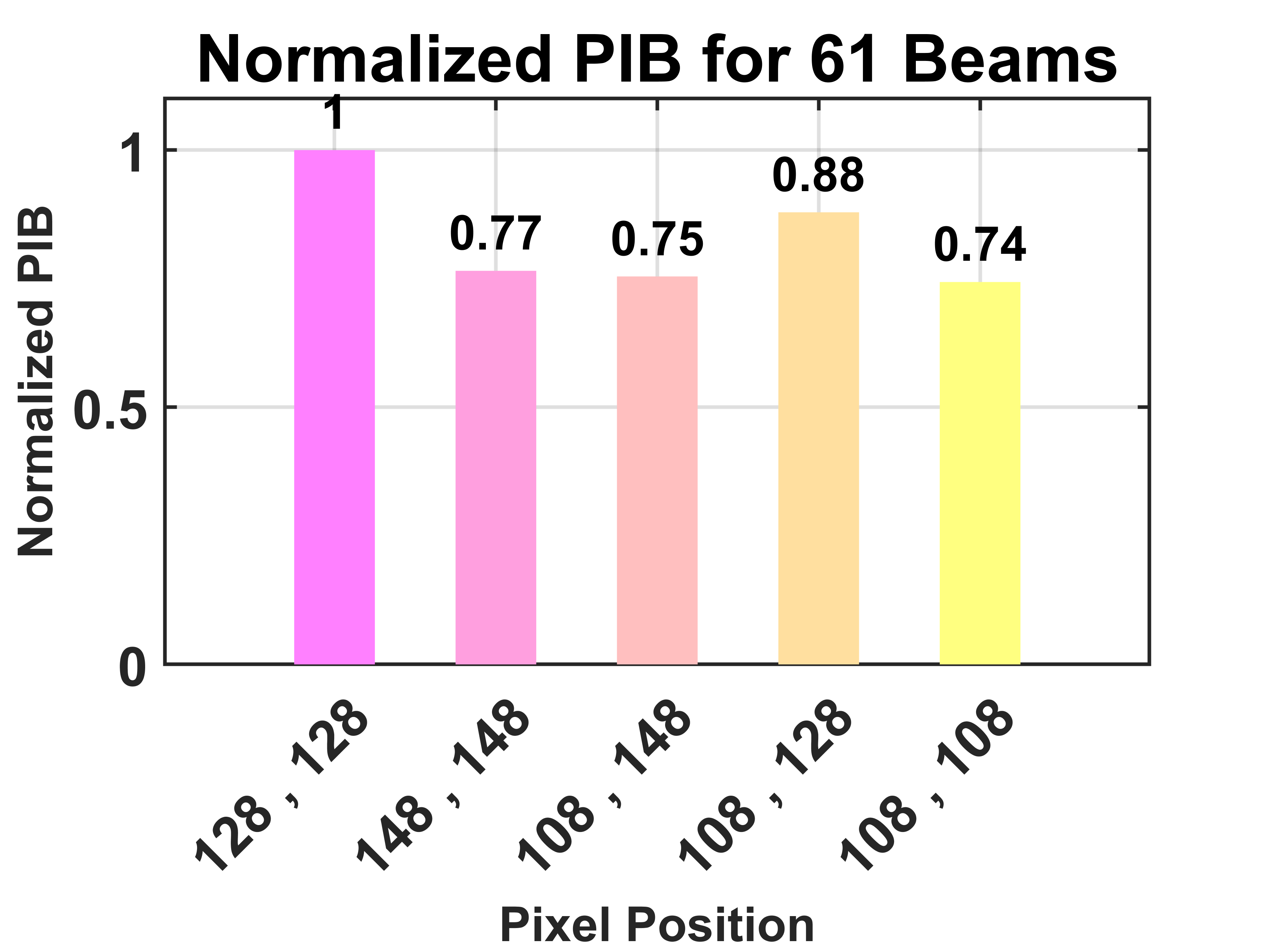}\label{fig:PIB_61}}

    \subfigure[]{\includegraphics[width=0.24\linewidth]{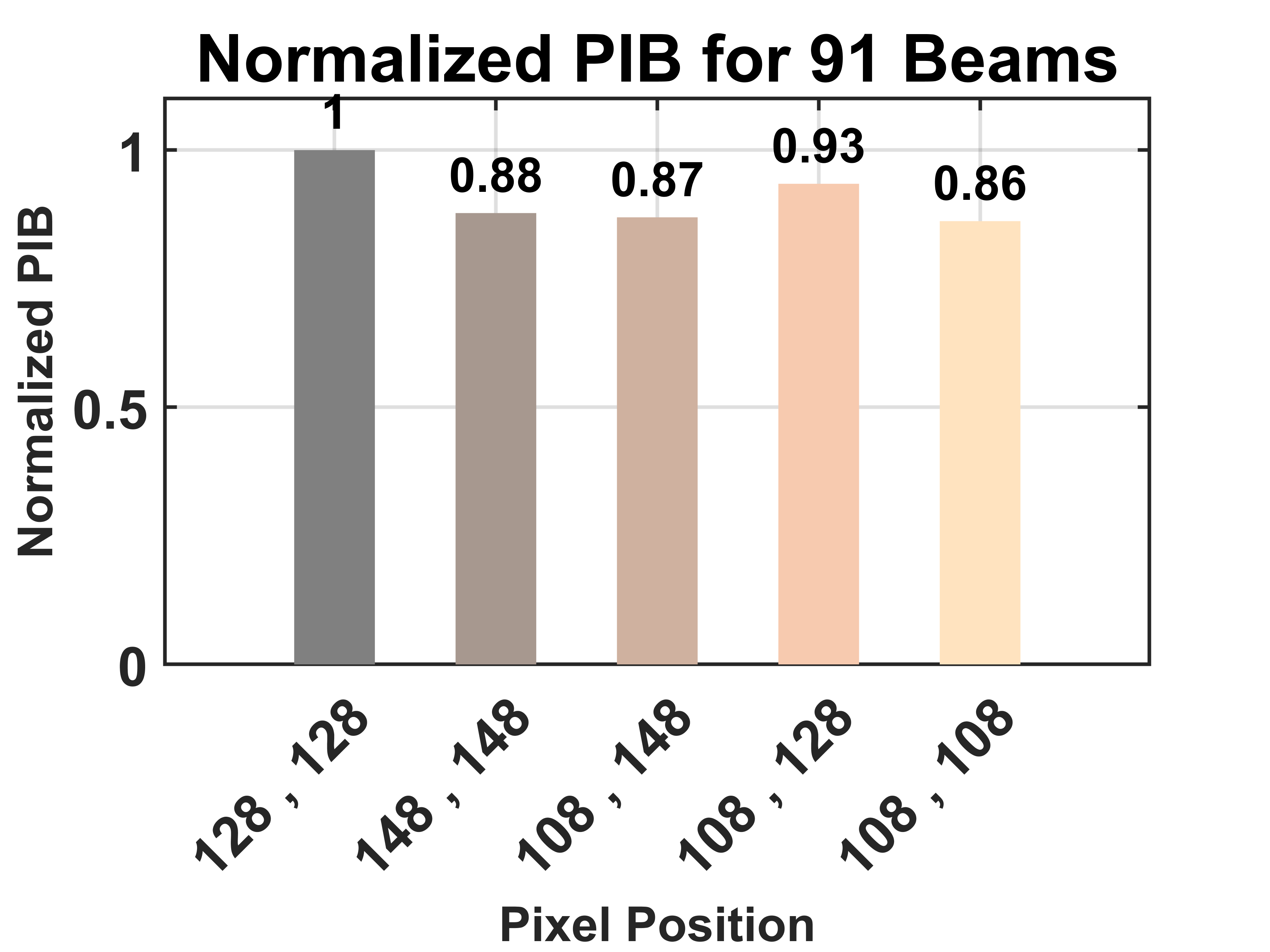}\label{fig:PIB_91}}
    \subfigure[]{\includegraphics[width=0.24\linewidth]{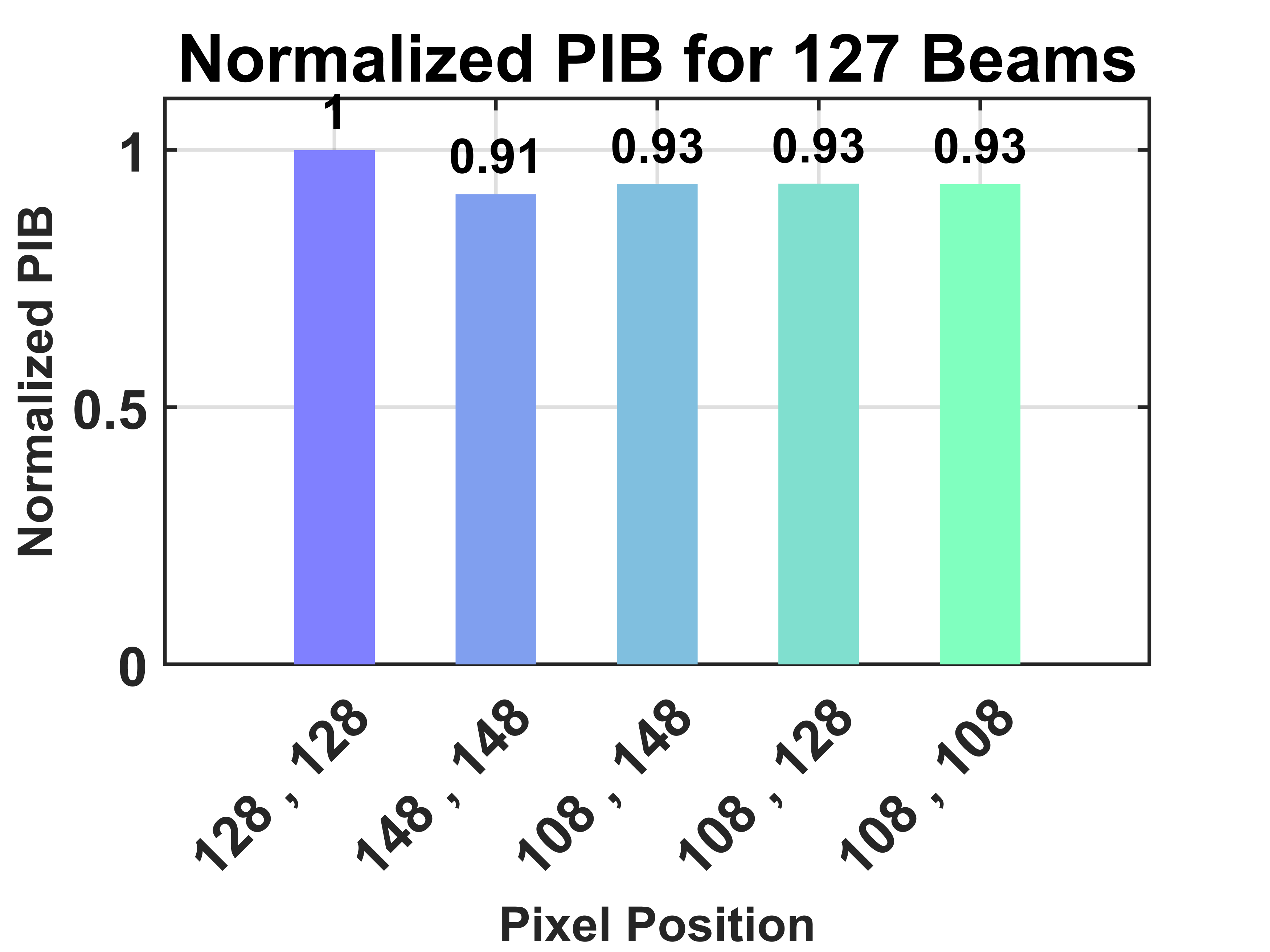}\label{fig:PIB_127}}
    \subfigure[]{\includegraphics[width=0.24\linewidth]{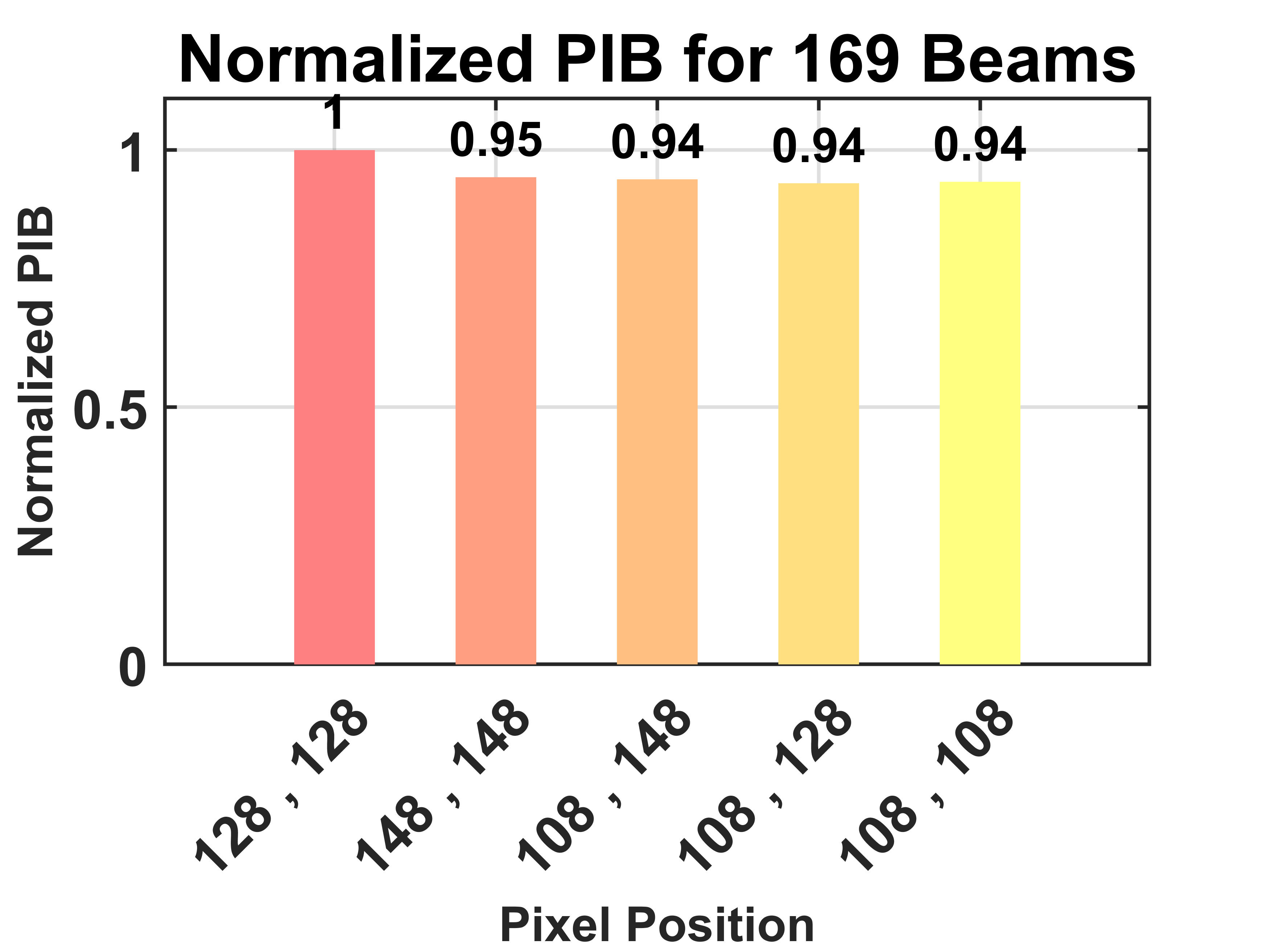}\label{fig:PIB_169}}
    \subfigure[]{\includegraphics[width=0.24\linewidth]{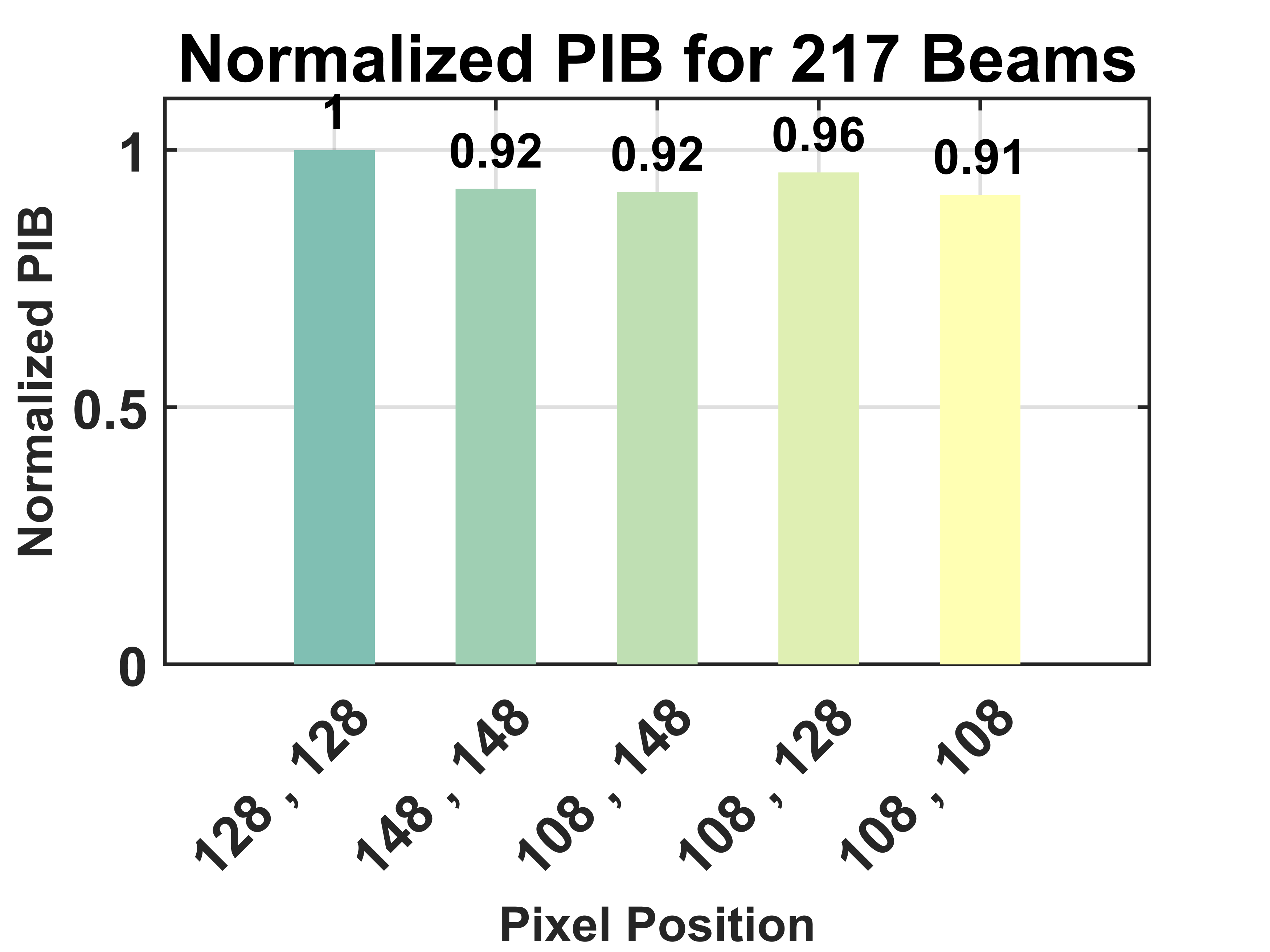}\label{fig:PIB_217}}

    \caption{CBC convergence analysis using Adagrad optimization at different pixel positions for varying numbers of beams: 
(a) 7 beams, (b) 19 beams, (c) 37 beams, (d) 61 beams, 
(e) 91 beams, (f) 127 beams, (g) 169 beams, and (h) 217 beams. 
The bar heights represent normalized PIB values for each pixel position.}

    \label{Beam_steer_plot}
\end{figure*}

\section{\textbf{Simulation}}

\subsection{\textbf{Numerical modeling of beam propagation}}
The simulation framework models a tiled-aperture coherent beam combining (TA-CBC) system, where $N$ Gaussian beams are arranged in a hexagonal geometry. The composite near-field distribution is expressed as
\begin{equation}
E(x,y,0) = \sum_{n=1}^{N} A(x-x_n,y-y_n)\,\exp\!\left[-i\phi_n\right],
\end{equation}
where $A(x,y)=\sqrt{\tfrac{2P}{\pi w_0^2}}\exp[-(x^2+y^2)/w_0^2]$ denotes the Gaussian envelope with waist $w_0$ and power $P$, $(x_n,y_n)$ are the sub-aperture centers, and $\phi_n$ represents the controllable phase of the $n$-th element~\cite{thomas2024physics}.  

Beam propagation from the near field to the far field is calculated using the angular spectrum method~\cite{matsushima2009band}, which numerically models free-space diffraction. The far-field distribution is obtained by applying a Fourier transform to the initial field and multiplying by the propagation transfer function $H(u,v,z)$. A band-limited filter is used to avoid aliasing and ensure numerical stability.

\subsection{\textbf{Beam combining evaluation methodology}}
The versatility of the framework allows evaluation under different operational scenarios. A general target mask $M(x,y)$ is defined in the far field to represent the desired region, which can correspond to a focused spot (steering) or an arbitrary profile (shaping).  

The performance is quantified using the power-in-the-target region (PITR) metric:
\[
\text{PITR} = \frac{\sum |E(x,y)|^2 M(x,y)}{\sum |E(x,y)|^2},
\]
PITR measures the fraction of energy delivered into the target region. Depending on the application, $M(x,y)$ can be chosen as a circular aperture (beam steering) or a structured mask (beam shaping).  

This generalized model highlights the flexibility of the TA-CBC platform: by only adjusting the phase distribution $\{\phi_n\}$ and the definition of $M(x,y)$, the same system can be applied to diverse requirements such as precise beam steering, static profile generation, or dynamic reconfiguration.


\subsection{\textbf{Numerical simulation setup}}
The numerical simulations are carried out on a computational window of \(256 \times 256\) pixels with a physical length scale in millimeters. Each beam in the array is modeled as a Gaussian mode with a waist radius of \( w_0 = 5 \, \text{mm} \), separated by \(12 \, \text{mm}\), and operating at a wavelength of \( \lambda = 1064 \, \text{nm} \). Since TA-CBC performance is evaluated in the far field, beam propagation is simulated accordingly. To demonstrate scalability, we extend our analysis from small arrays to ensembles of up to 217 beams, highlighting the generality, scalability, and robustness of the proposed approach.

\section {\textbf{Algorithm—Adagrad Optimizer}}

Phase optimization plays a central role in coherent beam combining because the far-field pattern is highly sensitive to the phase values assigned to each channel. Even small phase errors can significantly reduce combining efficiency or distort the desired beam shape, making reliable and stable convergence essential. Traditional gradient-descent methods often struggle in this environment, as a fixed global learning rate may produce slow updates for some phases while causing overshoot in others. Adaptive optimization is the key enabler for the phase control. Among the tested methods, a deep-learning-inspired Adagrad optimization approach demonstrated the fastest and most stable convergence, ensuring scalability and  robustness~\cite{thomas2024physics}.

An Adagrad optimizer that automatically adjusts the learning rate for each phase based on its accumulated gradient history~\cite{duchi2011adaptive}. Phases that experience large or frequent gradients receive smaller steps, improving stability, while less responsive phases are updated more aggressively, improving overall convergence speed. This parameter-wise adaptation makes Adagrad well suited for phase correction tasks, where different beams exhibit varying sensitivities, and it reduces the amount of manual tuning required for effective optimization.

The update rule is:
\[
\phi_i = \phi_{i-1} + \frac{\alpha}{\sqrt{G_i} + \epsilon} \, g_i,
\quad
G_i = G_{i-1} + g_i^2
\]
where $g_i$ is the gradient of the merit function $J(\phi)$ with respect to $\phi$ at iteration $i$, $\alpha$ is the base learning rate, and $\epsilon$ ensures numerical stability.

In this study, we used $\alpha=0.8$ and $\epsilon=10^{-10}$. This adaptive adjustment improves stability and convergence speed, making Adagrad an effective optimizer for both beam steering and beam shaping tasks in CBC~\cite{thomas2024physics}.


\section{\textbf{Laser beam engineering through CBC}}

\subsection{\textbf{Beam Steering and dynamic beam
shaping via sequential steering}}
 
Beam steering is the controlled redirection of optical energy to specific locations in the far-field~\cite{soni2025adaptive}. In many laser-based applications, the ability to steer a beam dynamically is essential, as it enables flexible and precise energy delivery without requiring physical motion of optical components~\cite{9893325}. Traditional approaches to steering, such as galvanometric scanners and mirror-based systems~\cite{bremer2025mathematical}, rely on mechanical movement. While effective, they are constrained by inertia, limited response speed, and long-term system fatigue. More advanced non-mechanical methods, such as liquid crystal spatial light modulators~\cite{he2019liquid} and acousto-optic deflectors, provide higher speeds but are limited in scalability and power handling.

In contrast, CBC provides a robust and scalable platform for beam steering through all-optical control. By adjusting the relative phases of multiple sub-aperture beams, the interference pattern in the far-field can be engineered to form a high-intensity spot at any arbitrary location as shown in figure\ref{Beam_steer_at_different_location}. This eliminates the need for moving parts, allowing ultrafast operation, and makes the method inherently suitable for high-power laser systems.

The steering process begins by evaluating the far-field field distribution $E_{\text{out}}(x,y)$ through angular spectrum propagation. A desired target location $(x_s,y_s)$ in the observation plane is selected, represented on the discrete pixel grid as
\begin{equation}
x_s = x_0 + p_x \Delta x, \quad y_s = y_0 + p_y \Delta y,
\end{equation}
where $(x_0,y_0)$ defines the observation plane center, $(p_x,p_y)$ are pixel indices, and $\Delta x, \Delta y$ are the sampling intervals.

To evaluate the steering performance, we define the power-in-the-bucket (PIB) metric, which quantifies the fraction of total optical power concentrated within a circular region of radius
\begin{equation}
r_{\text{PIB}} = \frac{1.22  \lambda z}{L_n},
\end{equation}
where $\lambda$ is the operating wavelength, $z$ is the propagation distance, and $L_n = 2(N_L a + R_a)$ is the effective aperture size determined by the number of beams $N_L$, the beam spacing $a$, and the beam radius $R_a$. The PIB is computed as
\begin{equation}
\text{PIB}(x_s,y_s) = \frac{\displaystyle \iint_{r \leq r_{\text{PIB}}} |E_{\text{out}}(x,y)|^2  dxdy}{\displaystyle \iint |E_{\text{out}}(x,y)|^2  dxdy},
\end{equation}
with $r = \sqrt{(x-x_s)^2+(y-y_s)^2}$ denoting the radial distance from the target. A higher PIB value corresponds to more effective beam focusing at the steered position, making it a reliable indicator of steering efficiency.

In our system, beam steering is achieved through adaptive phase control applied across all channels, enabling precise placement of the combined beam at target locations. The deep-learning-inspired Adagrad optimizer ensures stable steering even in large arrays, where maintaining phase alignment across many elements is essential.
The effectiveness of this scheme is validated by analyzing normalized PIB comparisons at different steering positions for varying array sizes (7, 19, 37, 61, 91, 127, 169, and 217 beams) and multiple target positions as shown in subfigures~\ref{fig:PIB_7}--\ref{fig:PIB_217} respectively.

As illustrated in figure \ref{Beam_steer_plot}, central steering locations yield the highest PIB values, while off-axis steering achieves slightly reduced but still significant focusing performance. The difference arises from the physics and optics of the CBC aperture: at the center position, constructive interference from all sub-apertures is maximized symmetrically, concentrating the largest fraction of total power into the focal spot. When steering to off-axis positions, part of the energy spreads into side lobes because of the finite aperture size and less uniform overlap, which reduces the peak intensity at the steered spot. 
\begin{figure*}[h!]
    \centering
    \subfigure[]   {\includegraphics[width=0.26\linewidth]{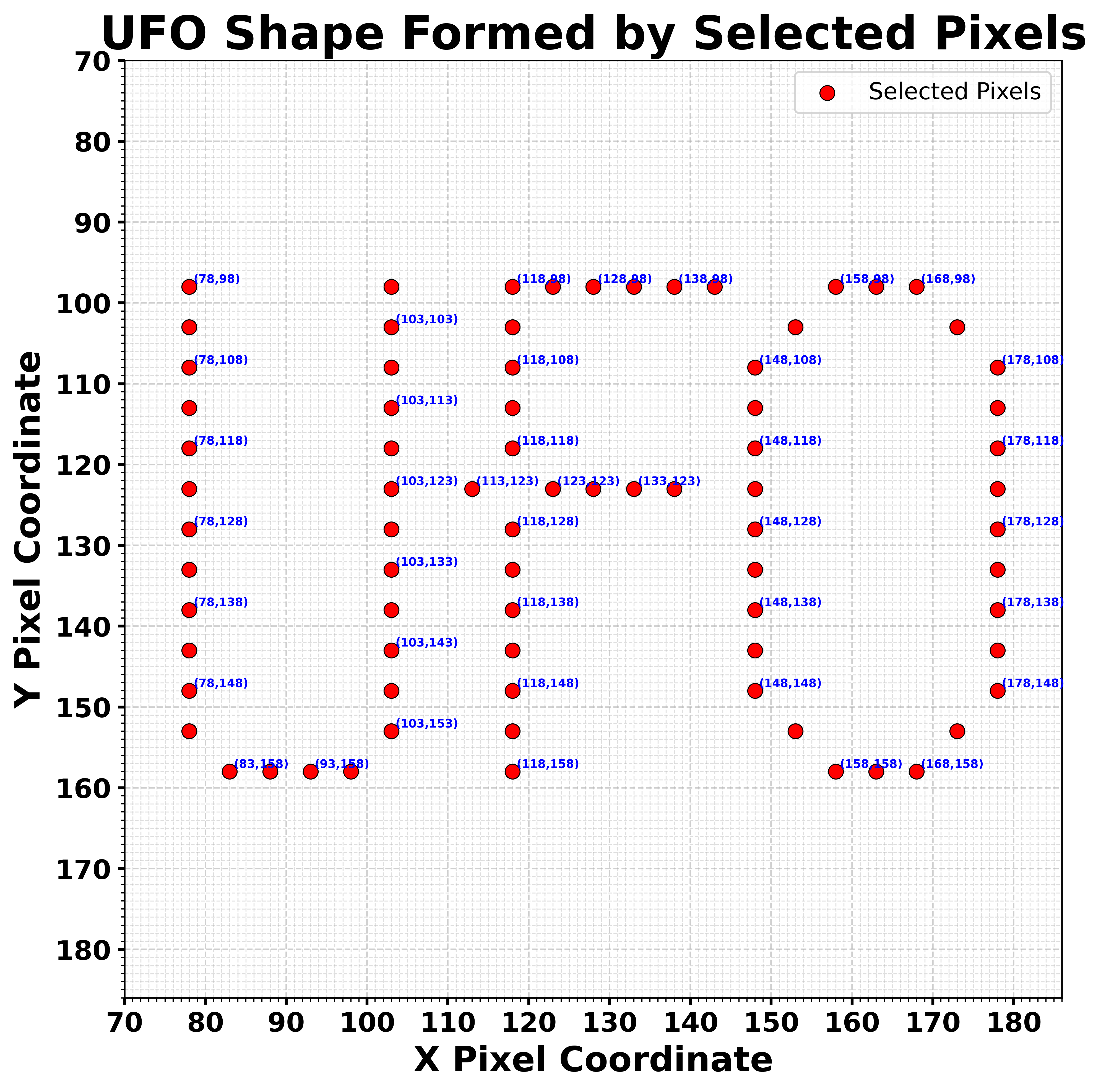}\label{fig:UFO_1}}
    \subfigure[]{\includegraphics[width=0.29\linewidth]{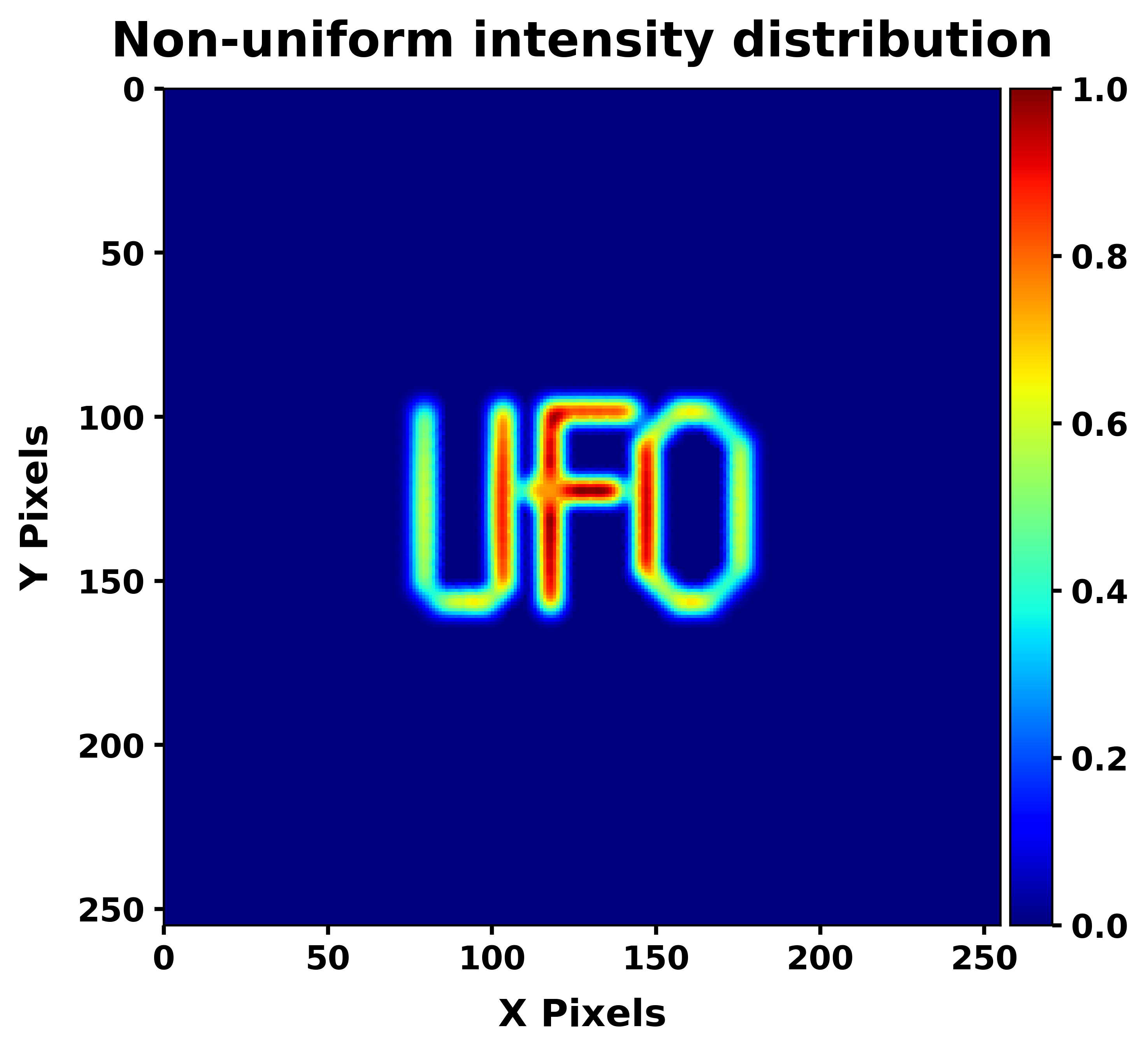}\label{fig:UFO_2}}
    \subfigure[]{\includegraphics[width=0.29\linewidth]{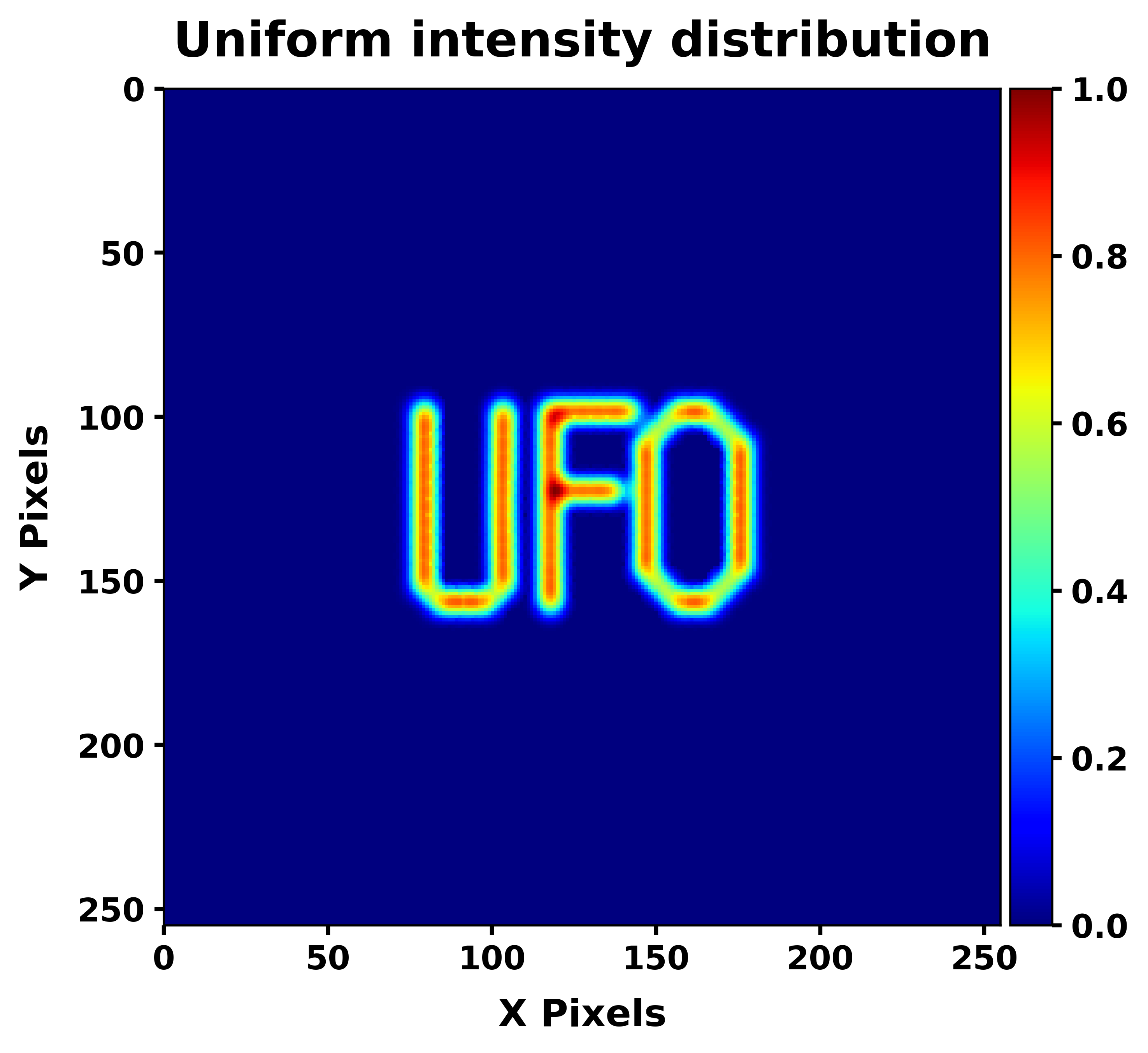}\label{fig:UFO_3}}

    \par\medskip 

    \subfigure[]{\includegraphics[width=0.90\linewidth]{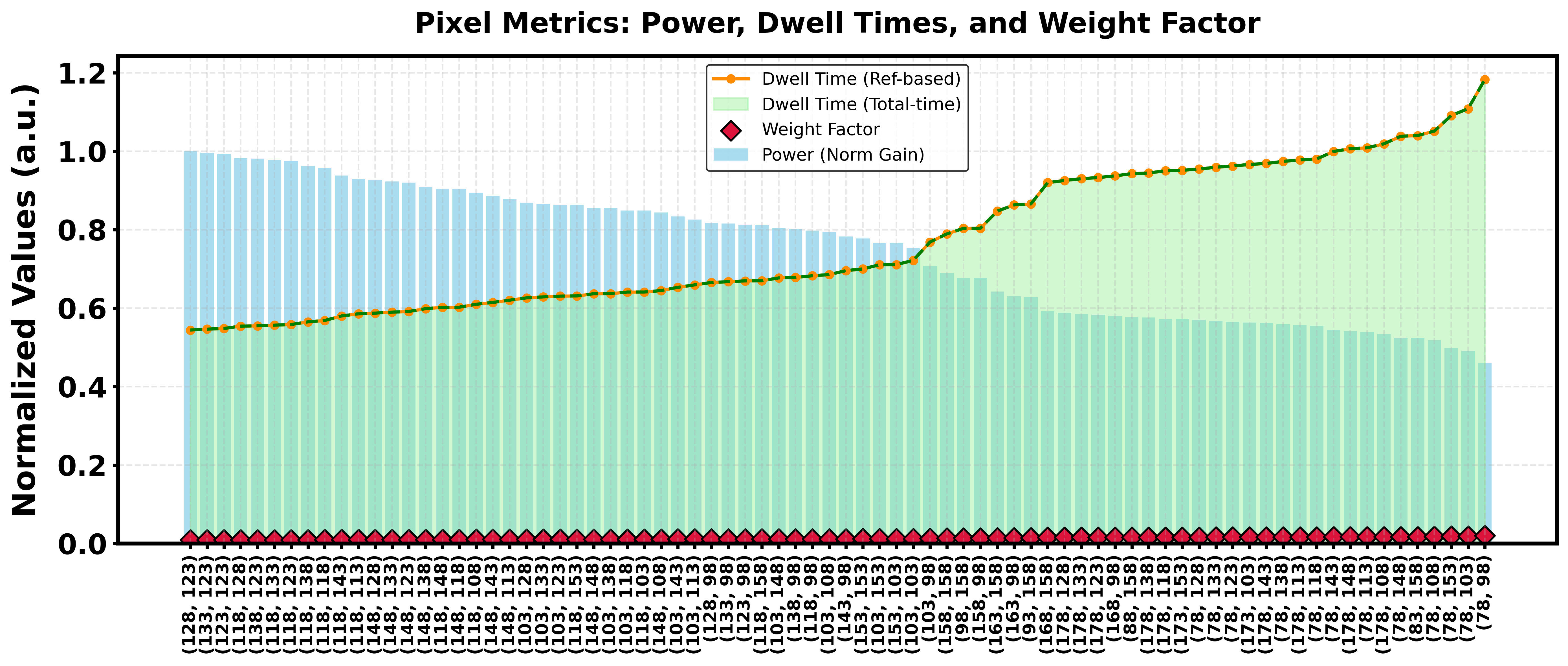}\label{fig:UFO_4}}
    


    \caption{\centering Formation and evaluation of the far-field “UFO” beam shape using sequential beam steering (a) Pixel grid showing the target “UFO” shape defined by selected beam positions.
    (b) Far-field intensity profile obtained with equal dwell time for all pixels, resulting in a non-uniform distribution.
    (c) Far-field intensity profile obtained with optimized (unequal) dwell times calculated from pixel gains, yielding nearly uniform intensity distribution across the “UFO” shape.
    (d) Quantitative analysis of pixel metrics: normalized gain, dwell time based on total time scanning method, dwell time based on weighted method, and corresponding weight factors for all selected pixels.}
    \label{Dwell_time_plot_with_Power}
\end{figure*}
\begin{figure*}[h!]
    \centering
    \subfigure[]   
    {\includegraphics[width=0.32\linewidth]{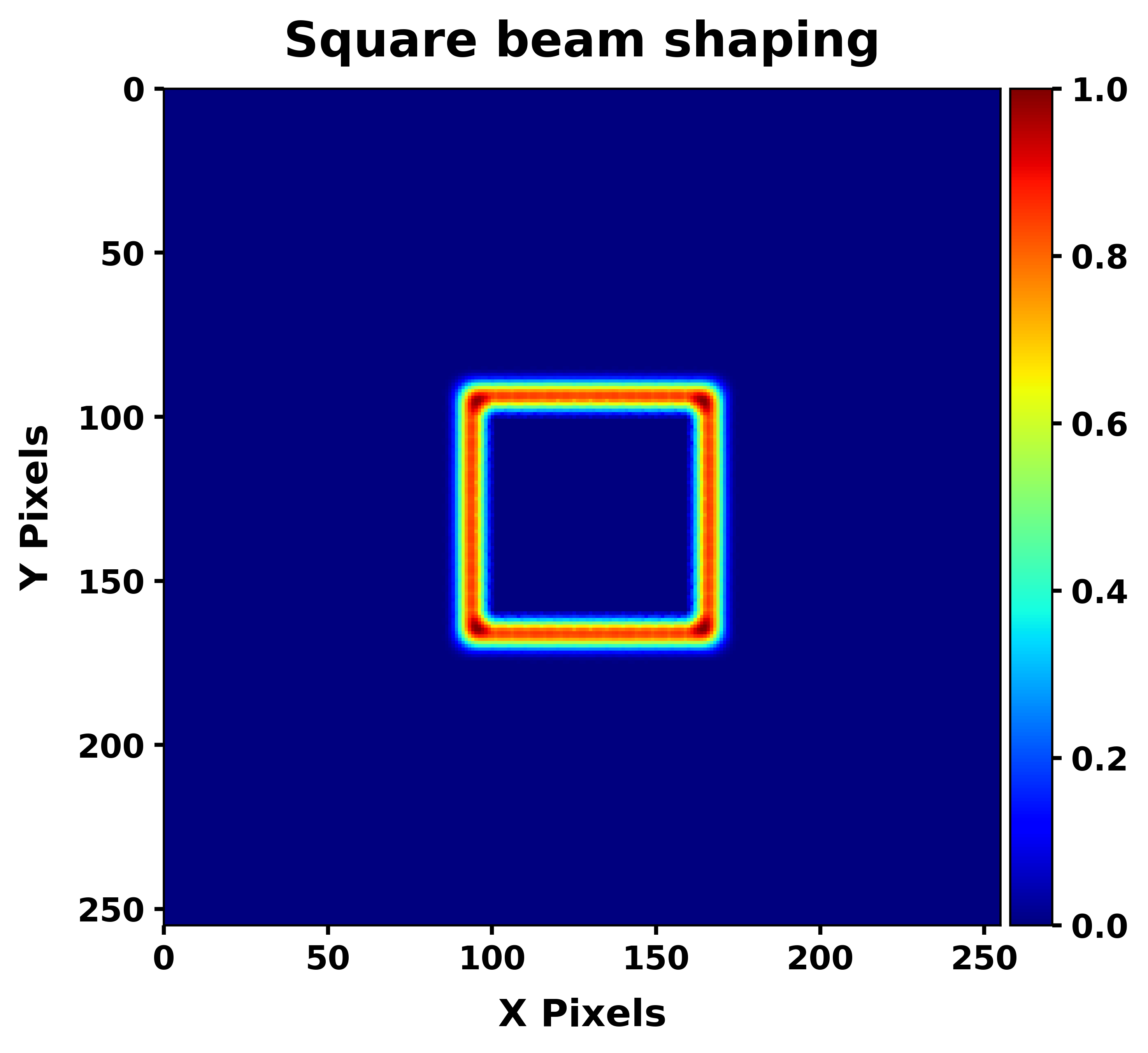}\label{fig:square_1}}
    \subfigure[]{\includegraphics[width=0.32\linewidth]{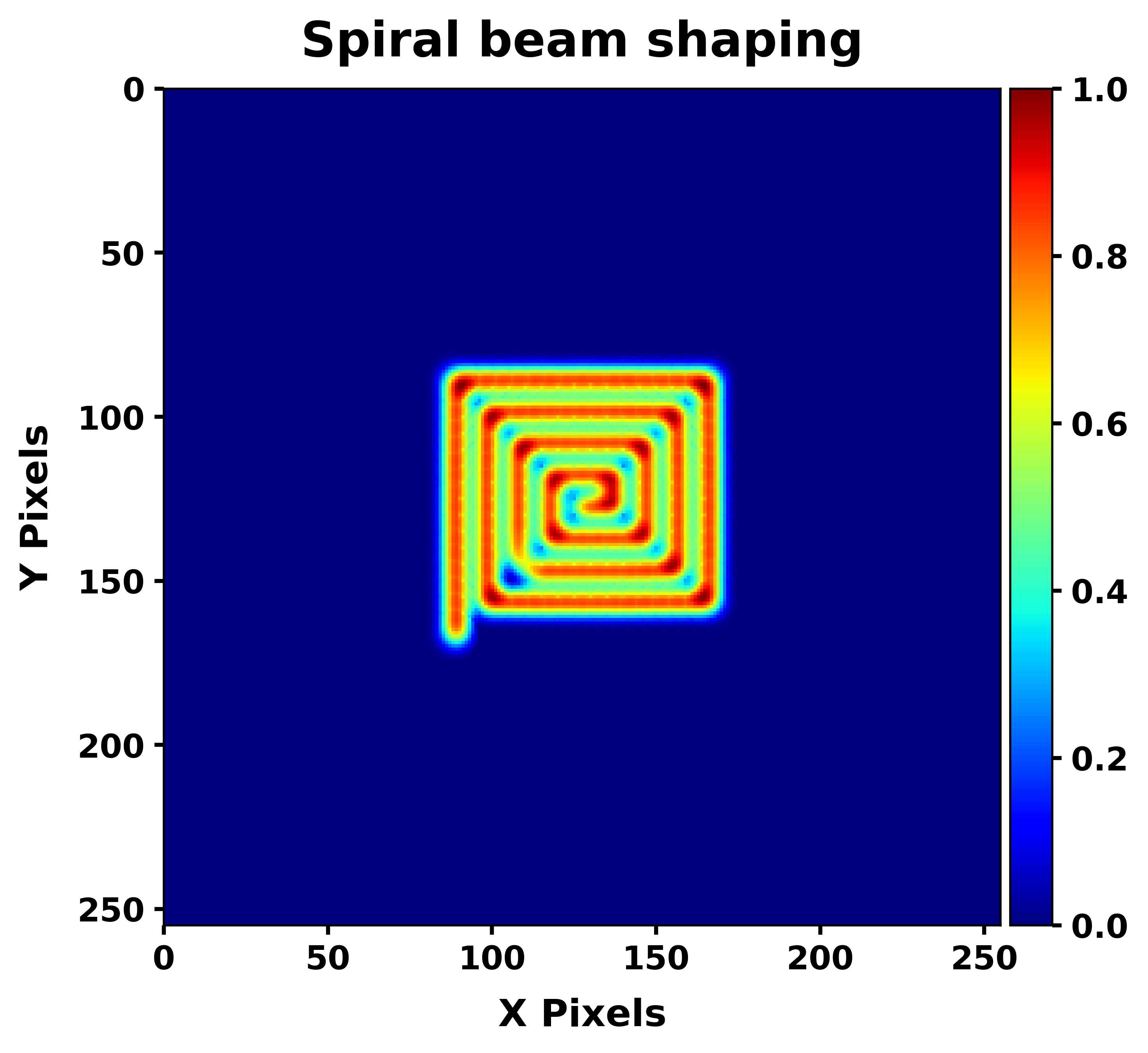}\label{fig:spiral_1}}
    \subfigure[]{\includegraphics[width=0.32\linewidth]{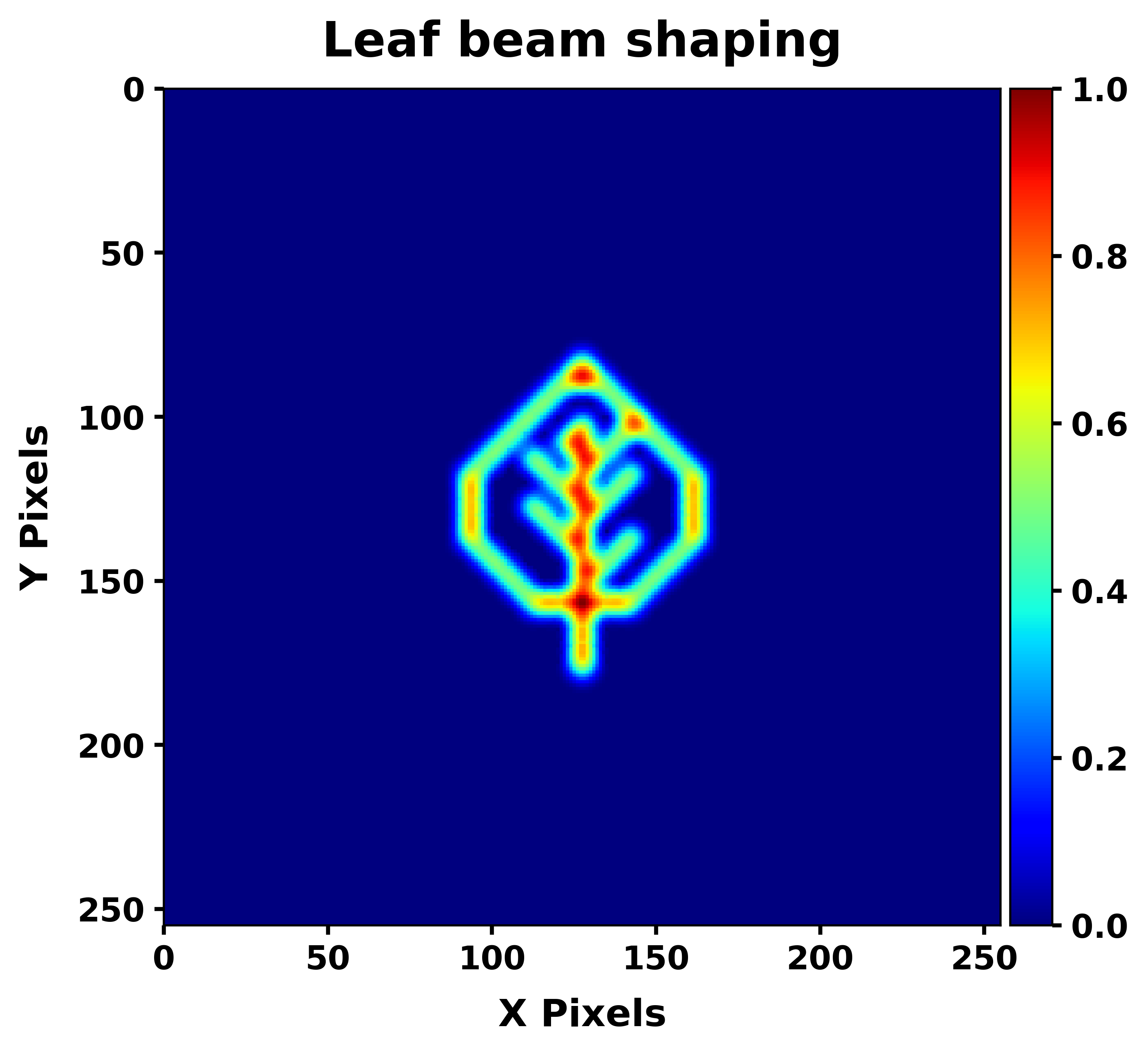}\label{fig:leaf_1}}

    \par\medskip 

    \subfigure[]{\includegraphics[width=0.32\linewidth]{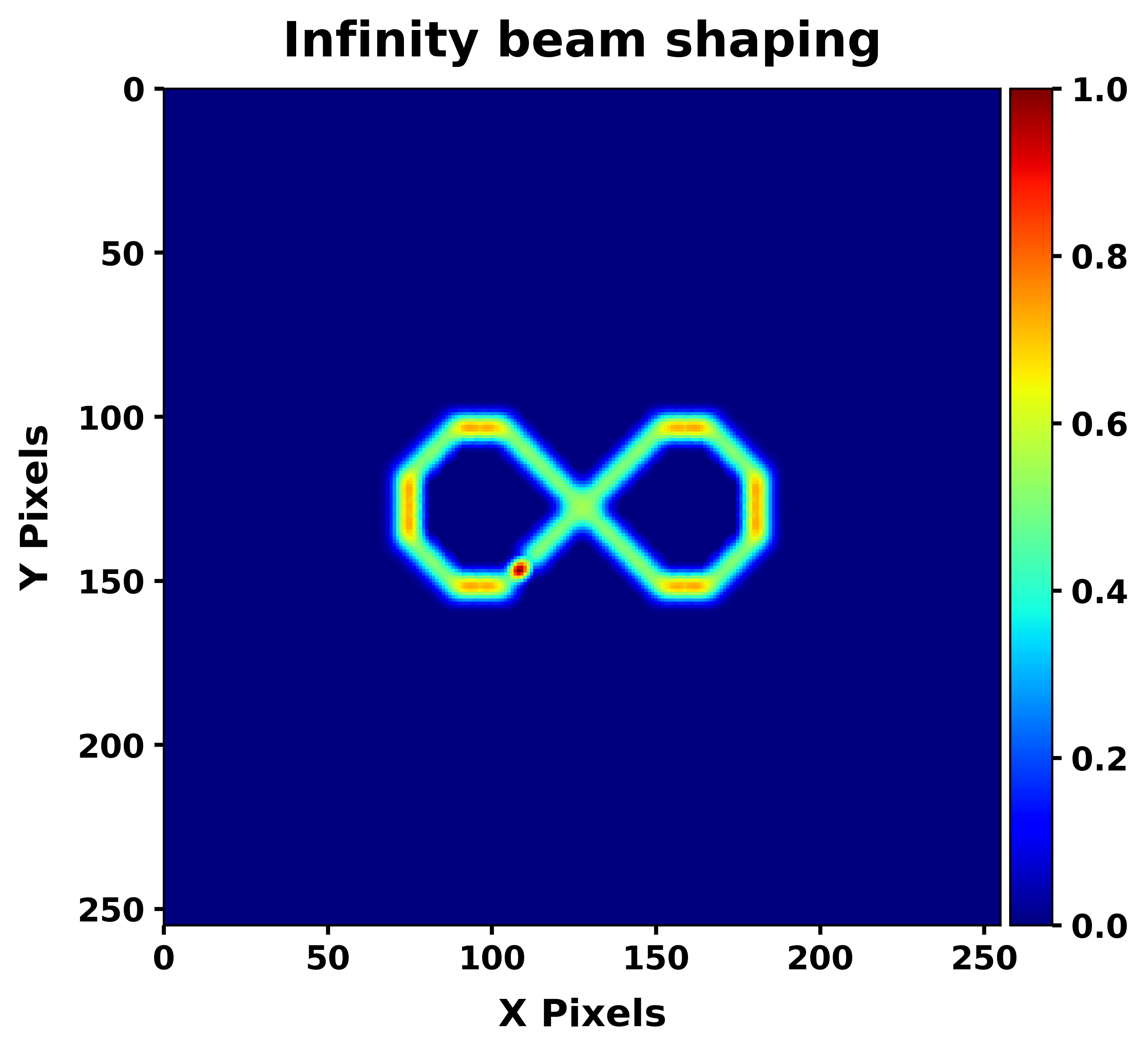}\label{fig:infinity_1}}
    \subfigure[]{\includegraphics[width=0.32\linewidth]{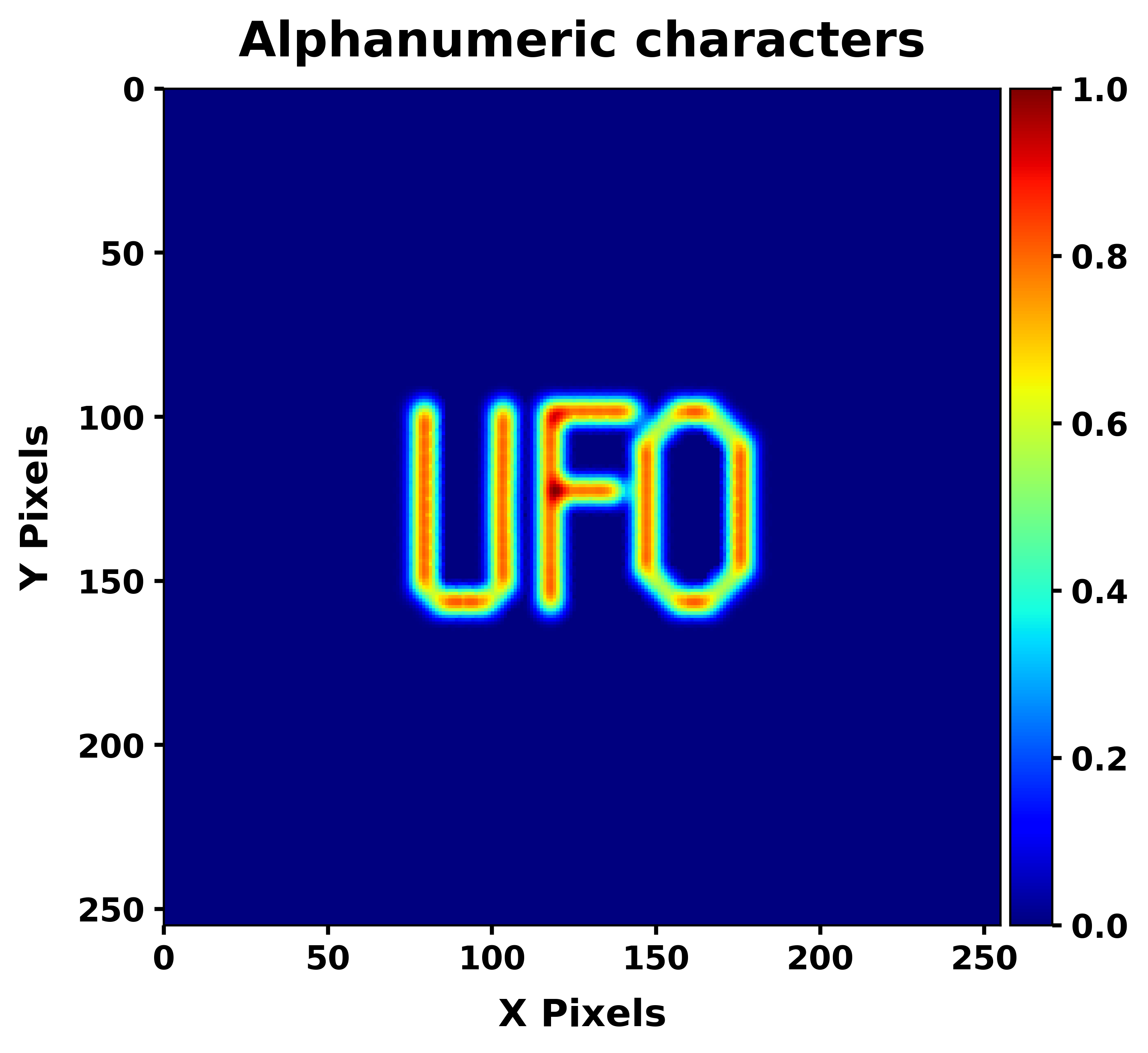}\label{fig:alphanumeric_1}}
    \subfigure[]{\includegraphics[width=0.32\linewidth]{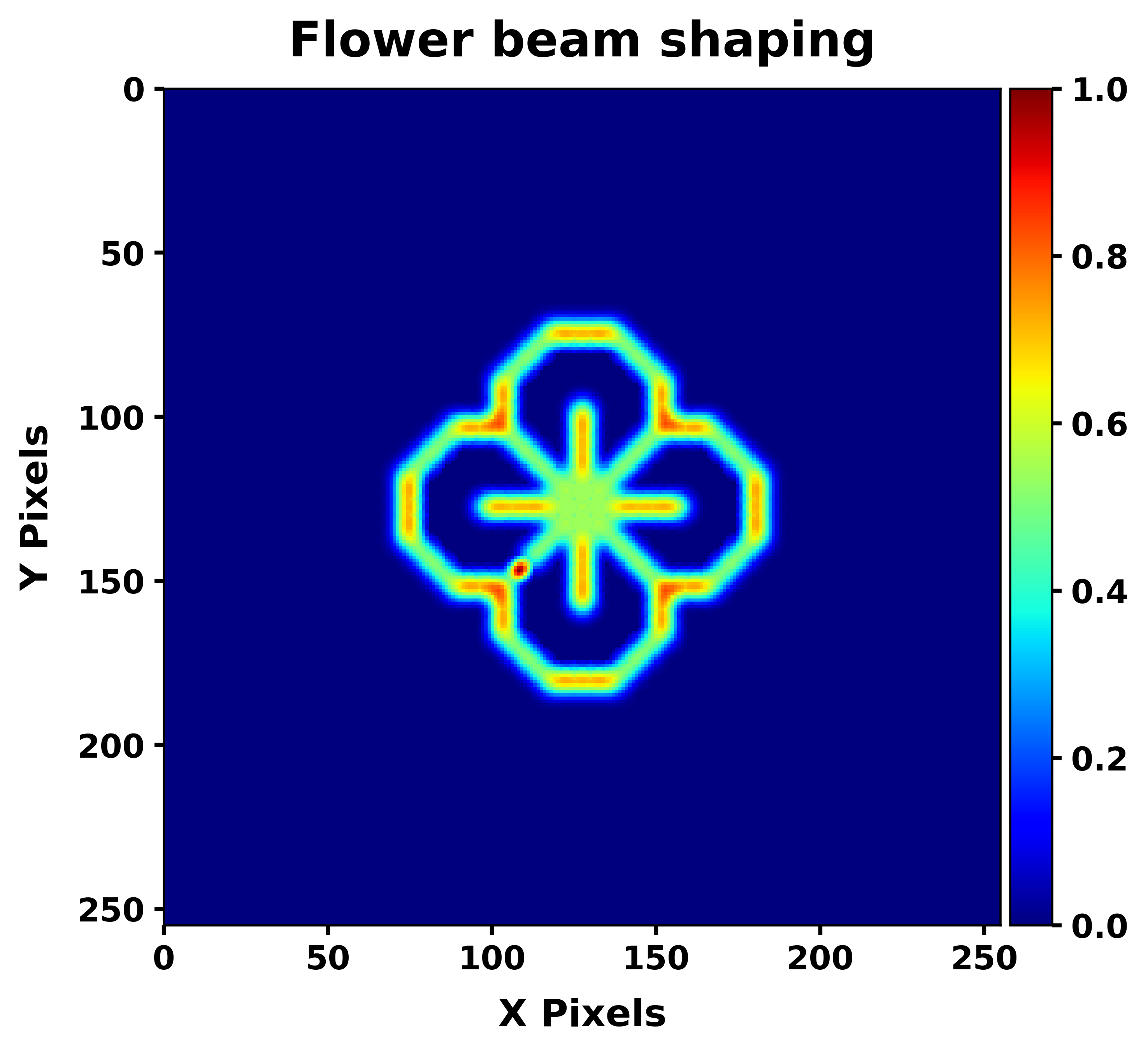}\label{fig:flower_1}}

    \caption{\centering Dynamic beam shaping through sequential steering using stored phase maps and the dwell-time concept to achieve near-equal power distribution across all target patterns.}
    \label{Beam_steer_different_shape}
\end{figure*}
However, our study indicates that increasing the number of channels can reduce the variation in power at different positions, as shown in figure~\ref{Beam_steer_plot}. For smaller arrays (e.g., 7 and 19 beams - subfigures ~\ref{fig:PIB_7} \& \ref{fig:PIB_19} ), the normalized PIB shows significant fluctuations across steering positions, indicating weaker stability and reduced efficiency in steering. As the number of beams increases (37 to 91 beams - subfigures ~\ref{fig:PIB_37}--\ref{fig:PIB_91}), the normalized PIB values become more uniform, with reduced variation across the selected positions, reflecting improved robustness of beam steering. Notably, at higher beam counts (127 to 217 beams - subfigures ~\ref{fig:PIB_127}--\ref{fig:PIB_217}), the normalized PIB values remain consistently close to unity across all positions, confirming that larger coherent arrays enhance steering accuracy and maintain high power concentration at the desired target pixel. These results confirm that increasing the number of beams not only improves the steering fidelity but also minimizes power loss during off-axis steering, with the 217-beam case offering the most stable and efficient performance among all tested configurations shows in subfigure~\ref{fig:PIB_217}.

Beyond simple single-point steering, the method is extended to sequential steering, which allows beam shaping by guiding the focus across multiple target positions in a controlled manner. In this case, the steering is not limited to a single static point but follows a predefined sequence of locations, effectively distributing energy across a pattern. This sequential steering enables the generation of arbitrary far-field intensity shapes. In our implementation, the phase values were stored only for the central region of the computational grid, enclosing the portion of the beam where a significant fraction of the total power is concentrated. The limits of this region vary with the total number of beams and the chosen grid size and are selected such that the intensity gradually decreases to about 50\% of the normalized peak at the beam center. This ensures that the shaping process operates on the most power-dense area of the field, where the profile exhibits a natural rise toward the center and a smooth decay outward, providing an effective envelope for beam engineering. This selection also avoids storing redundant pixels in the low-power tails, which contribute little to the effective beam formation. Based on this criterion, a grid was defined in both directions and the corresponding phase maps were stored. For sequential steering, these stored phase sets were applied according to the desired target shape, enabling controlled beam shaping through adaptive phase selection. The resulting far-field intensity distribution can then be expressed as the superposition of these sequentially steered states:

\begin{equation}
I_{\text{shape}}(x,y) = \sum_{k=1}^{M} w_k \, |E_{\text{out}}^{(k)}(x,y)|^2,
\end{equation}

where $M$ is the number of steering steps, $E_{\text{out}}^{(k)}(x,y)$ is the far-field distribution at the $k^{\text{th}}$ steering position, and $w_k$ is a weighting factor that controls the contribution of each step. This approach allows programmable beam shaping using only phase adjustments, without requiring additional hardware elements.  

However, when shaping more complex structures, an intrinsic challenge arises due to diffraction. In the far-field domain, constructive interference is strongest near the center, while higher spatial frequencies decay rapidly, resulting in an uneven power distribution: central pixels inherently carry more energy due to the physics of Fourier propagation, where most of the optical power concentrates at low spatial frequencies (near the center), while higher frequencies decay outward. If a target shape is reconstructed directly from these stored phase maps, this imbalance manifests as non-uniformity, with central regions appearing brighter and outer regions weaker as shown in subfigure~\ref{fig:UFO_2}.  

To mitigate this, we introduce a simple but effective correction strategy based on a time–power relation. The key idea is to compensate intensity differences dynamically: pixels with naturally lower intensity are held for longer durations, while pixels with higher intensity are held for shorter durations. 



In practice, this relation can be implemented in two complementary ways.

\subsubsection*{\textbf{1. Reference-based scaling}}
A high-intensity pixel (e.g., the one corresponding to maximum normalized power) is chosen as the reference. The dwell time for any other pixel $k$ is then scaled as  

\begin{equation}
T_k = T_{\text{ref}} \cdot \frac{G_{\text{ref}}}{G_k},
\end{equation}

where $G_k$ is the measured intensity (or gain) at pixel $k$, and $G_{\text{ref}}$ is the intensity at the reference pixel. The reference pixel corresponds to the maximum normalized power of the combined beam, while the pixels at the edges of the selected central region represent approximately 50\% of this maximum value. Consequently, the dwell times at these peripheral pixels are proportionally larger than at the reference, reflecting their lower intensity.

Since the effective beam shape is expressed as a weighted superposition of steering states, these dwell times directly determine the weighting coefficients used for beam shaping:  

\begin{equation}
w_k = \frac{T_k}{\sum_j T_j}.
\end{equation}

\subsubsection*{\textbf{2. Total-time normalization}}
If a fixed overall scan duration $T_{\text{tot}}$ is imposed, the dwell times are distributed according to normalized inverse gains:  

\begin{equation}
r_k = \frac{1}{G_k}, \quad 
T_k = \frac{r_k}{\sum_j r_j} \, T_{\text{tot}},
\end{equation}

where $r_k$ is the inverse-gain factor for the $k^{\text{th}}$ pixel and the denominator $\sum_j r_j$ ensures proper normalization across all pixels $j = 1, \dots, M$.  

In this case as well, the corresponding weight is simply  

\begin{equation}
w_k = \frac{T_k}{T_{\text{tot}}} = \frac{r_k}{\sum_j r_j},
\end{equation}

which shows that the weighting factors $w_k$ in the beam-shaping superposition naturally emerge from the dwell-time allocation strategy.

Both formulations, though expressed differently, are mathematically equivalent - they both yield dwell-time weights that equalize energy delivery across the target shape: an adaptive time allocation strategy that equalizes energy delivery across the beam shape. This correction reduces diffraction-induced non-uniformity and improves pattern contrast, leading to a more accurate far-field reconstruction without altering the stored phase information.

As an illustrative example, figure~\ref{Dwell_time_plot_with_Power} demonstrates the application of the proposed method for a 217-beam combination. In this case, the active region of the grid was selected between indices 73 and 183 along both the x and y directions, which corresponds to the central portion of the beam where the intensity falls to about 50\% of the normalized maximum at the edges. Subfigure~\ref{fig:UFO_1} shows the target “UFO” pattern defined by the selected pixels, while subfigure~\ref{fig:UFO_2} depicts the initial non-uniform intensity distribution obtained from direct superposition. After applying dwell-time–based weighting, subfigure~\ref{fig:UFO_3} shows the resulting uniform intensity profile across the shape. The subfigure~\ref{fig:UFO_4} summarizes the pixel-wise metrics: normalized power, dwell times (both reference-based and total), and the derived weighting factors. As expected, pixels at lower intensity levels exhibit proportionally larger dwell times to compensate for reduced power, leading to uniform energy delivery across the entire target region.

Following this demonstration, figure~\ref{Beam_steer_different_shape} presents additional examples of dynamic beam shaping achieved through sequential steering. It is important to note that the shapes shown in figure~\ref{Beam_steer_different_shape} are not fixed beams that exist all at once. Instead, they are formed by adding up the energy from many steering points. What we see in the figure is the overall energy distribution created when the beam moves through all those points in sequence. Different geometric and symbolic patterns—including a square, spiral, leaf, infinity symbol, alphanumeric text “UFO”, and a flower-like structure are successfully generated in subfigures~\ref{fig:square_1}-~\ref{fig:flower_1} respectively. These results highlight the versatility of the proposed method, where precomputed steering states are adaptively weighted to realize arbitrary target profiles. The uniformity across each shape confirms that the dwell-time–based weighting effectively compensates for intensity variations in the steering trajectories, resulting in stable and well-defined far-field distributions. Collectively, these examples demonstrate that the approach can flexibly engineer complex beam geometries while maintaining consistent energy delivery, illustrating its potential for adaptive laser beam engineering in diverse applications.

Overall, CBC-based non-mechanical beam steering, enhanced with deep learning optimization, provides a highly flexible platform that combines ultrafast response, robustness, and scalability. The ability to extend steering into sequential operation opens a pathway toward programmable beam shaping, making this approach particularly beneficial in industrial laser processing, additive manufacturing, free-space optical communication, and high-power directed-energy systems.

\subsection{\textbf{Adaptive static beam shaping}}

The spatial distribution of a laser beam plays a crucial role in many applications. Different beam profiles are often needed at various stages of a process, including preheating, welding, and post-treatment. Conventional approaches to achieving this flexibility rely on numerous optical elements or dynamic modulation systems, which can increase system complexity and introduce alignment challenges. Beam shaping addresses these limitations by enabling tailored intensity profiles that meet the specific needs of diverse applications~\cite{shekel2024dynamic,bakhtari2024review}. Whether for materials processing, high-resolution imaging, directed energy systems, or free-space optical communications, the ability to precisely control the far-field beam distribution significantly impacts system performance and utility. In many cases, a Gaussian-shaped beam is not optimal; instead, structured intensity profiles such as rectangles, rings, or triangles provide consistent energy deposition, improved targeting precision, or enhanced system interaction.  

In this work, we develop a generalized mask-based optimization framework for beam shaping in CBC systems. Multiple laser beams are phase-controlled to interfere constructively in the far field, producing the desired intensity pattern. Traditional approaches lose efficiency as the array size grows, restricting their use in large-scale systems. The presented framework addresses this challenge and provides a stable, scalable solution capable of generating a variety of high-quality beam profiles.

\begin{figure*}[h!]
    \centering
     \subfigure[]{\includegraphics[width=0.24\linewidth]{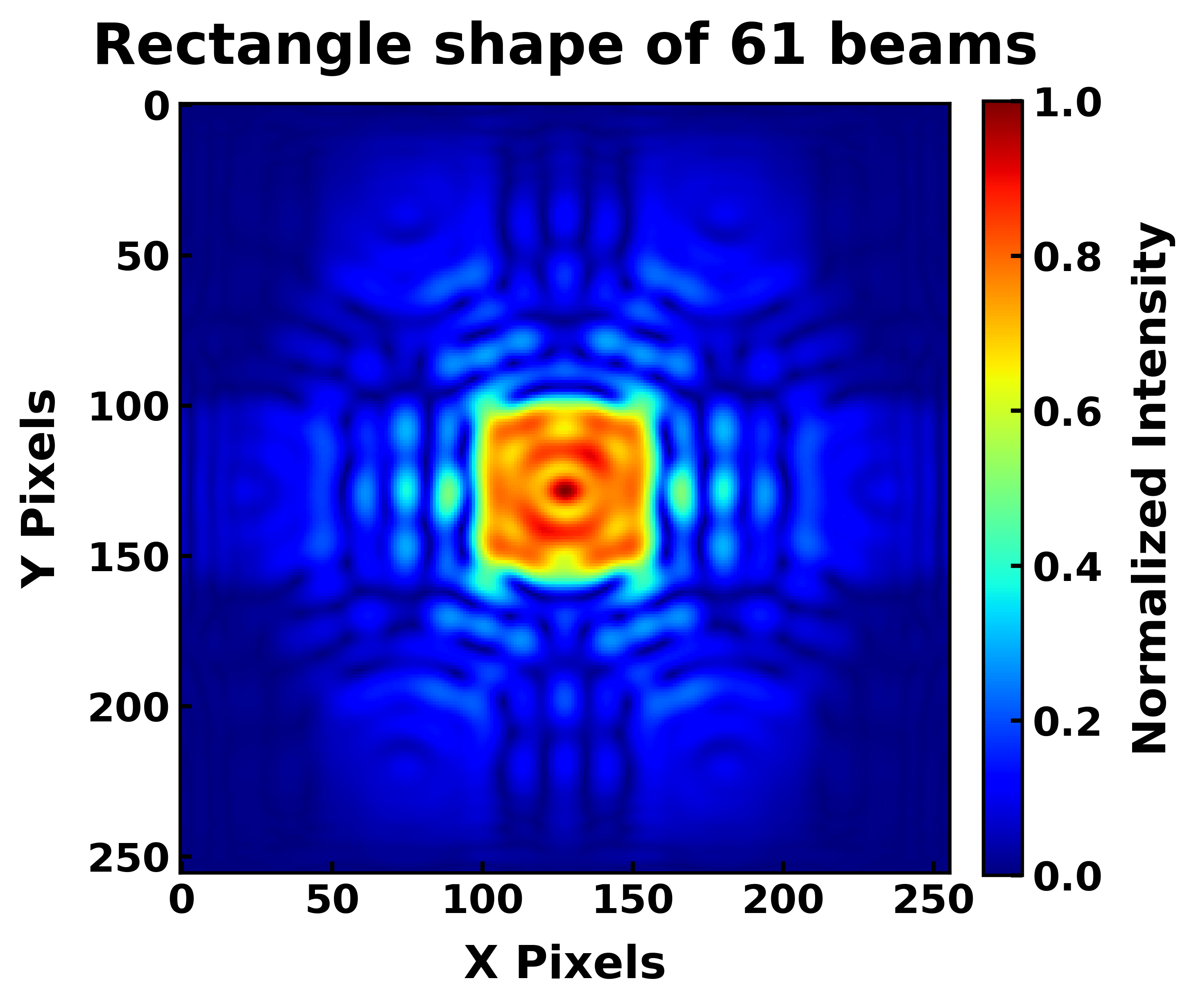}\label{fig:61_ractangle_1}}
    \subfigure[]{\includegraphics[width=0.24\linewidth]{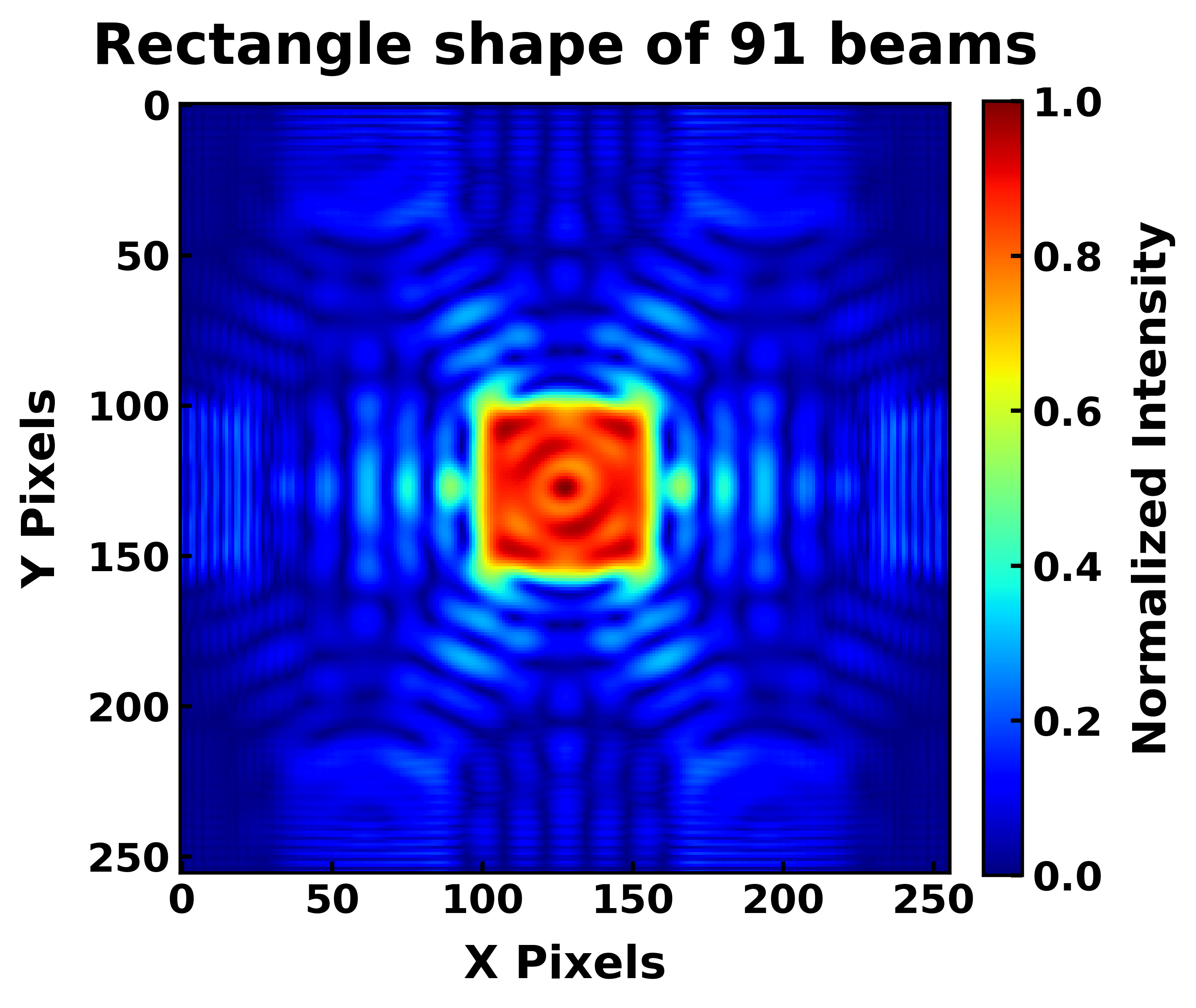}\label{fig:91_ractangle_1}}
    \subfigure[]{\includegraphics[width=0.24\linewidth]{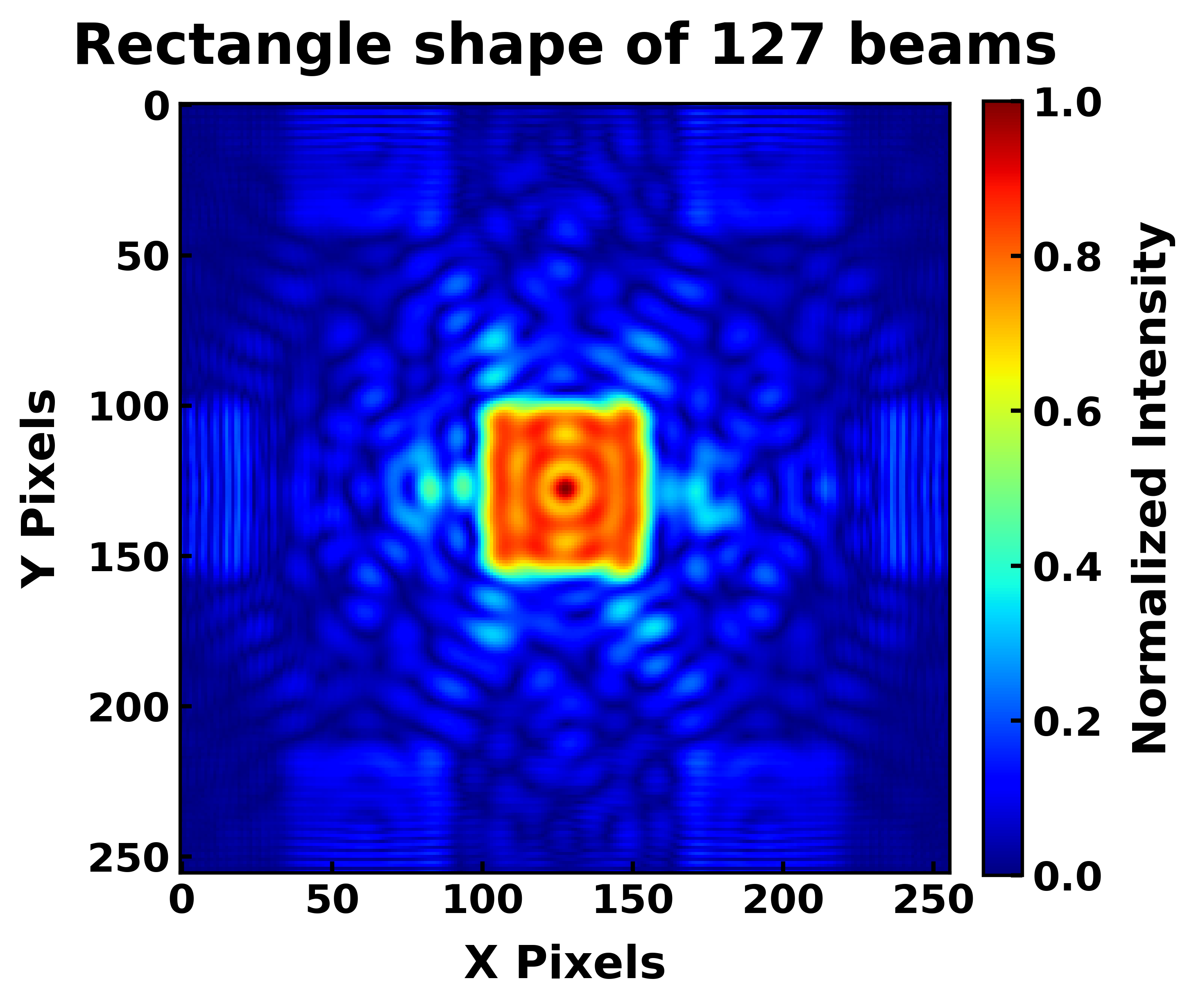}\label{fig:127_ractangle_1}}
    \subfigure[]{\includegraphics[width=0.24\linewidth]{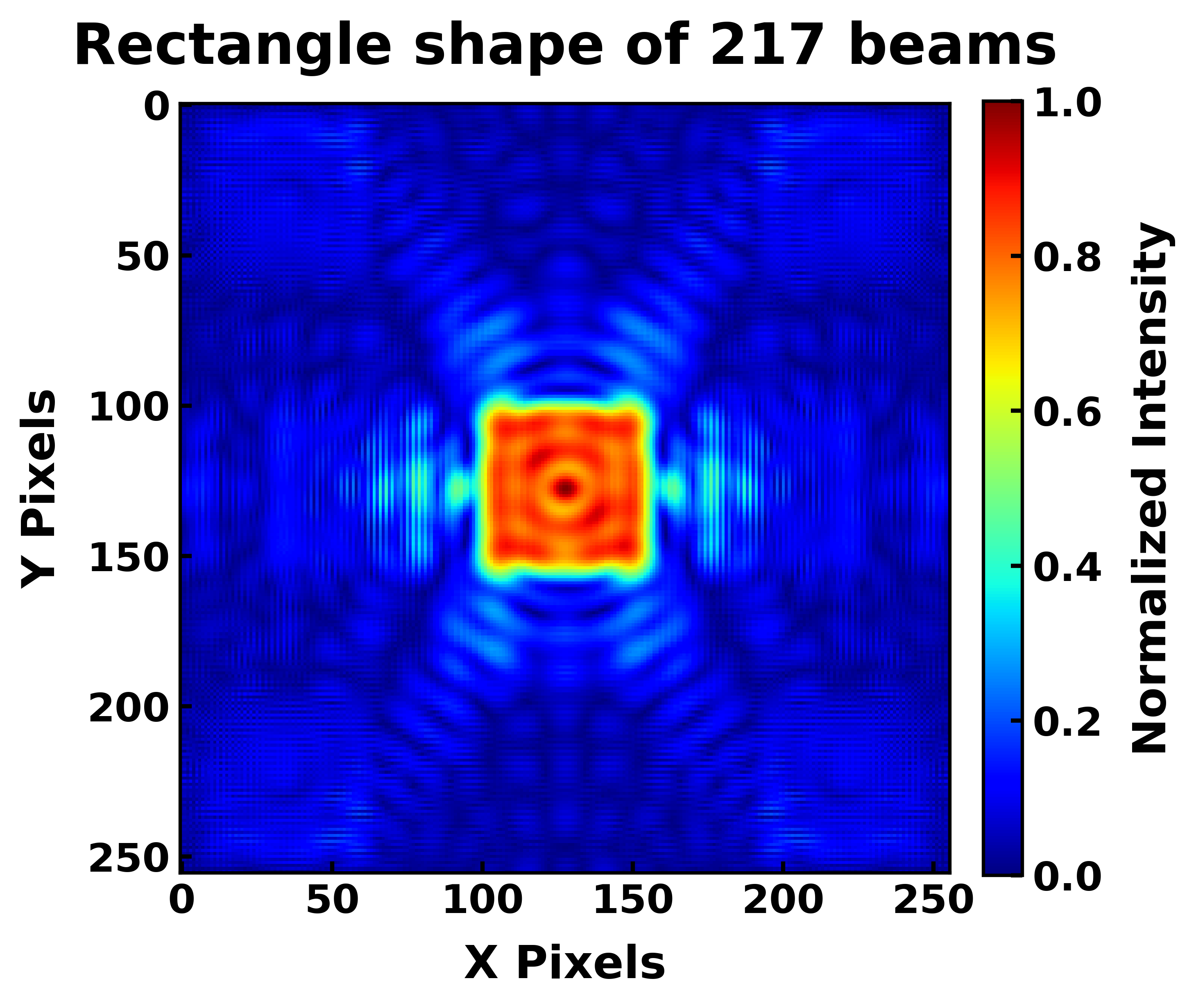}\label{fig:217_ractangle_1}}
    \par\medskip 

    \subfigure[]{\includegraphics[width=0.95\linewidth]{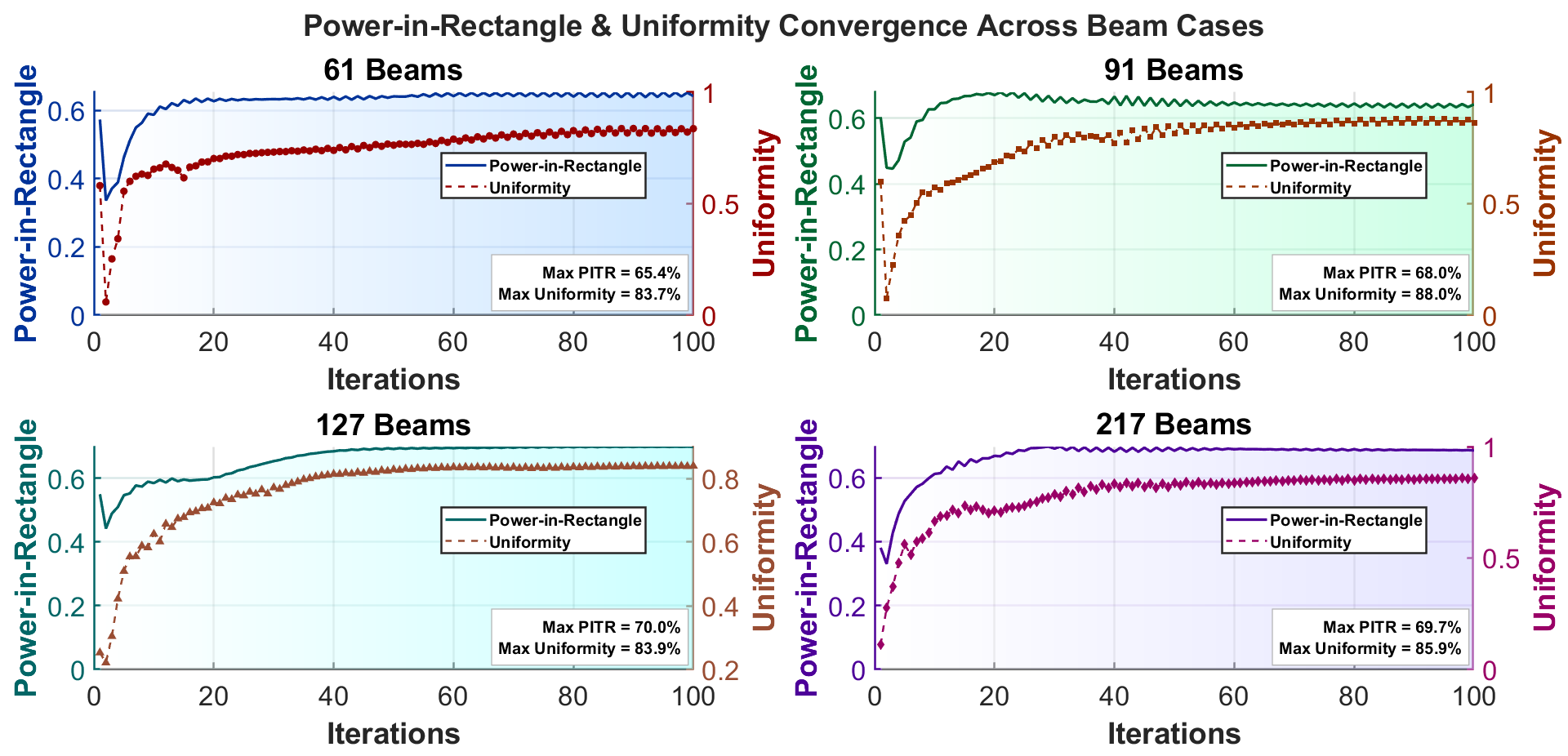}\label{fig:Ractangle_combined_4_different_beams_1}}

    \caption{\centering CBC convergence analysis using the Adagrad algorithm for rectangle shape: (a–d): Optimized far-field intensity profiles for 61, 91, 127, and 217 beams, (e): Convergence plots (iteration vs power-in-rectangle and uniformity) for different beams}
    \label{Square_plot}
\end{figure*}
\begin{figure*}[h!]
    \centering
     \subfigure[]{\includegraphics[width=0.24\linewidth]{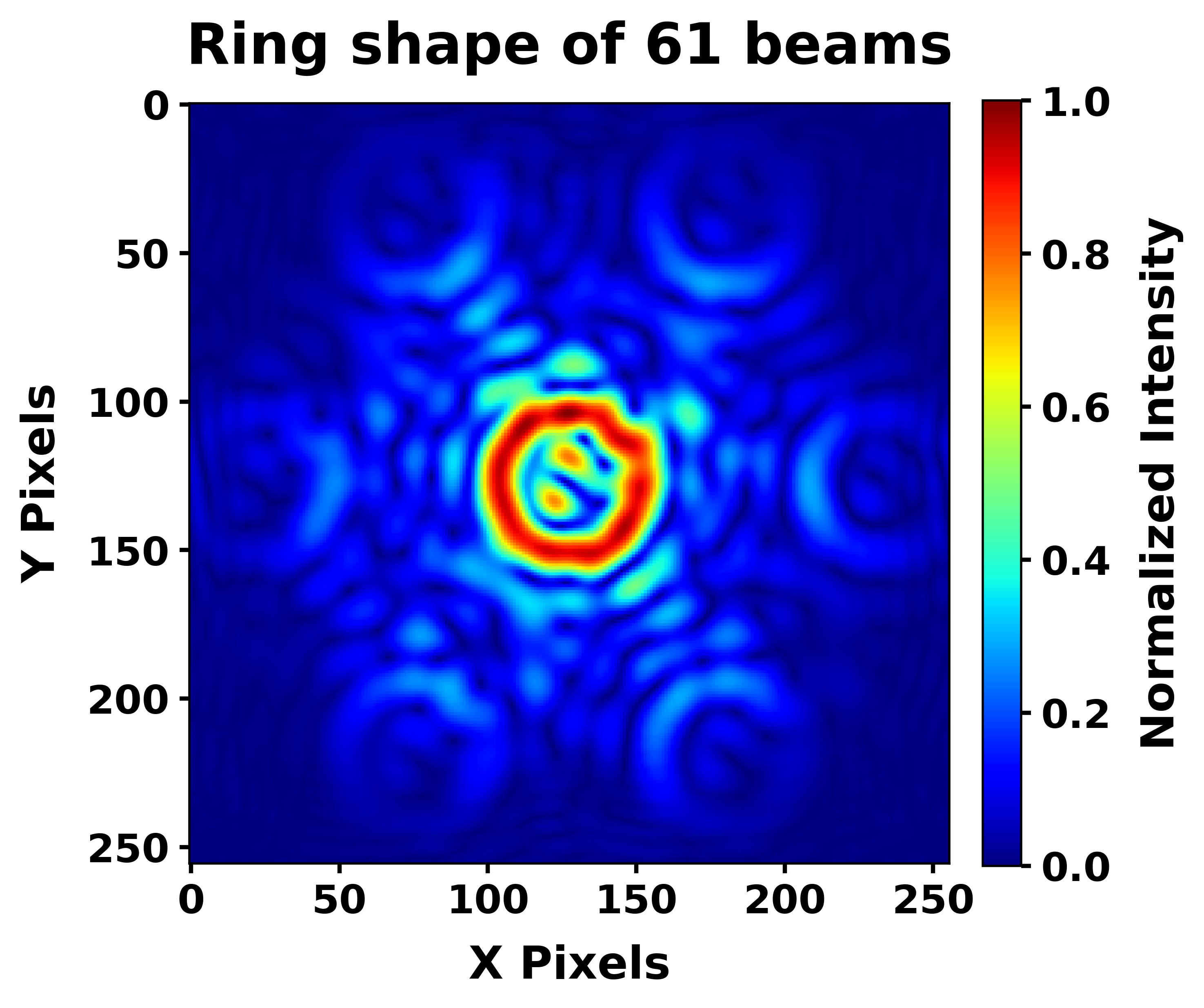}\label{fig:61_ring_1}}
    \subfigure[]{\includegraphics[width=0.24\linewidth]{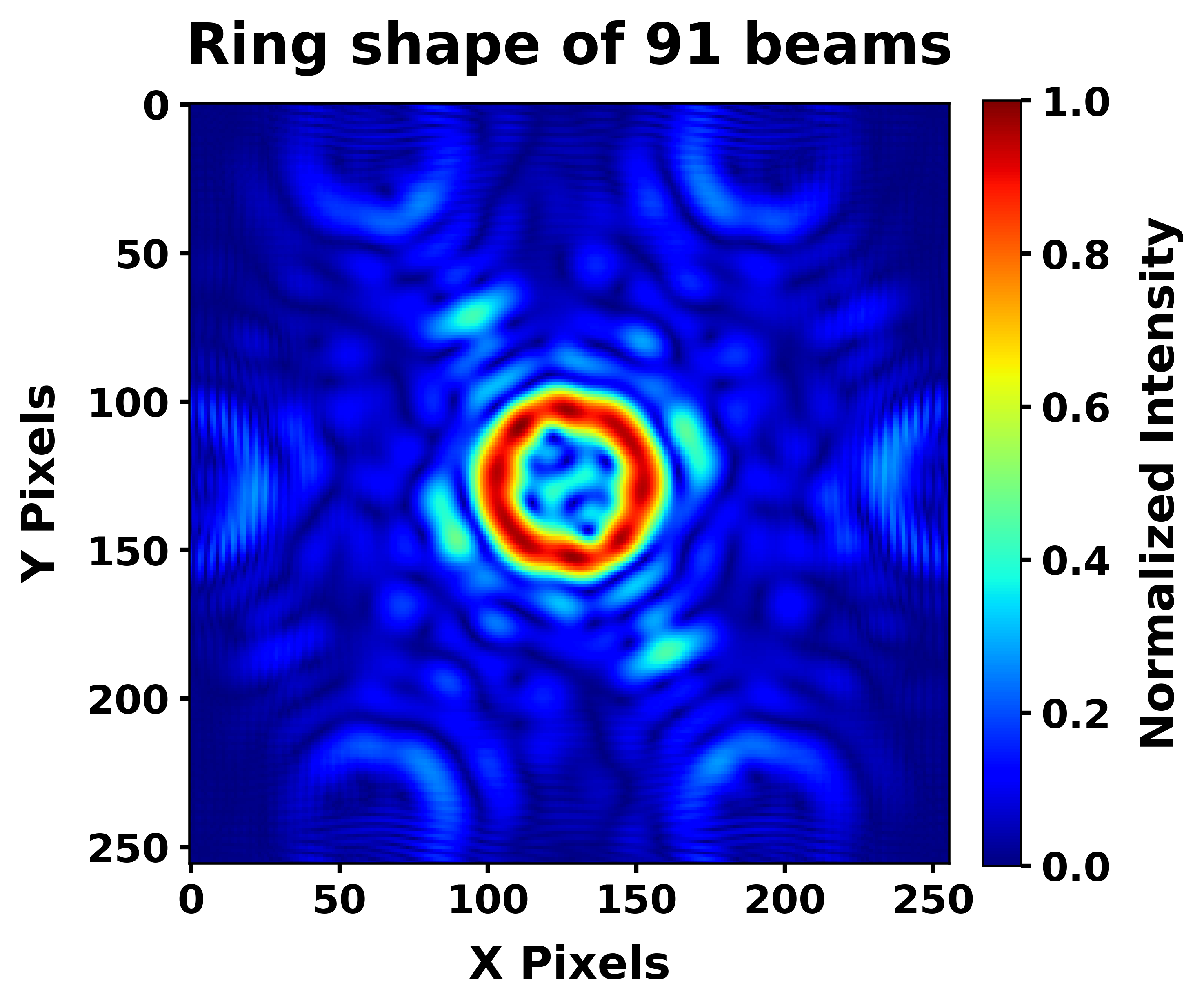}\label{fig:91_ring_1}}
    \subfigure[]{\includegraphics[width=0.24\linewidth]{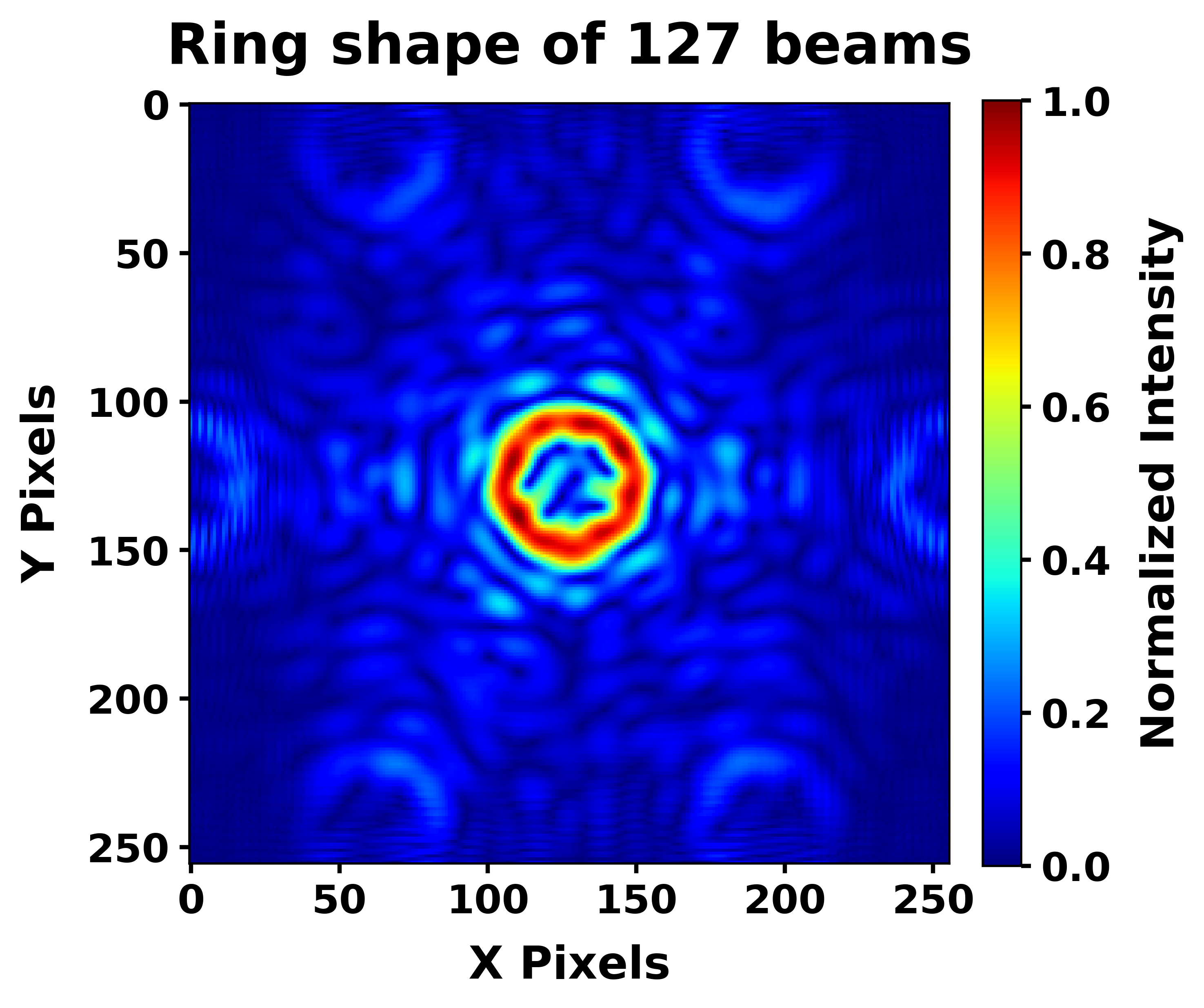}\label{fig:127_ring_1}}
    \subfigure[]{\includegraphics[width=0.24\linewidth]{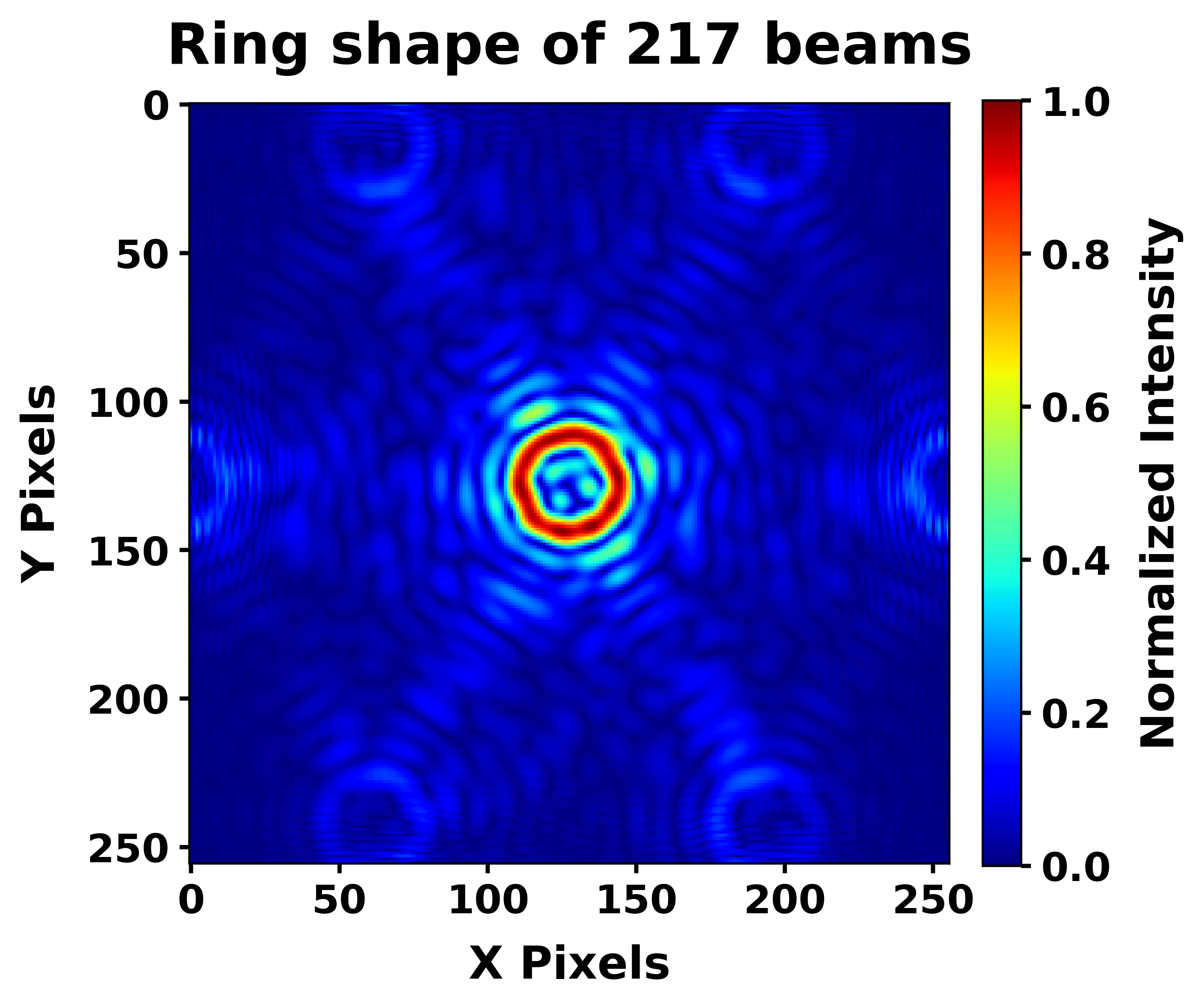}\label{fig:217_ring_1}}
    \par\medskip 

    \subfigure[]{\includegraphics[width=0.95\linewidth]{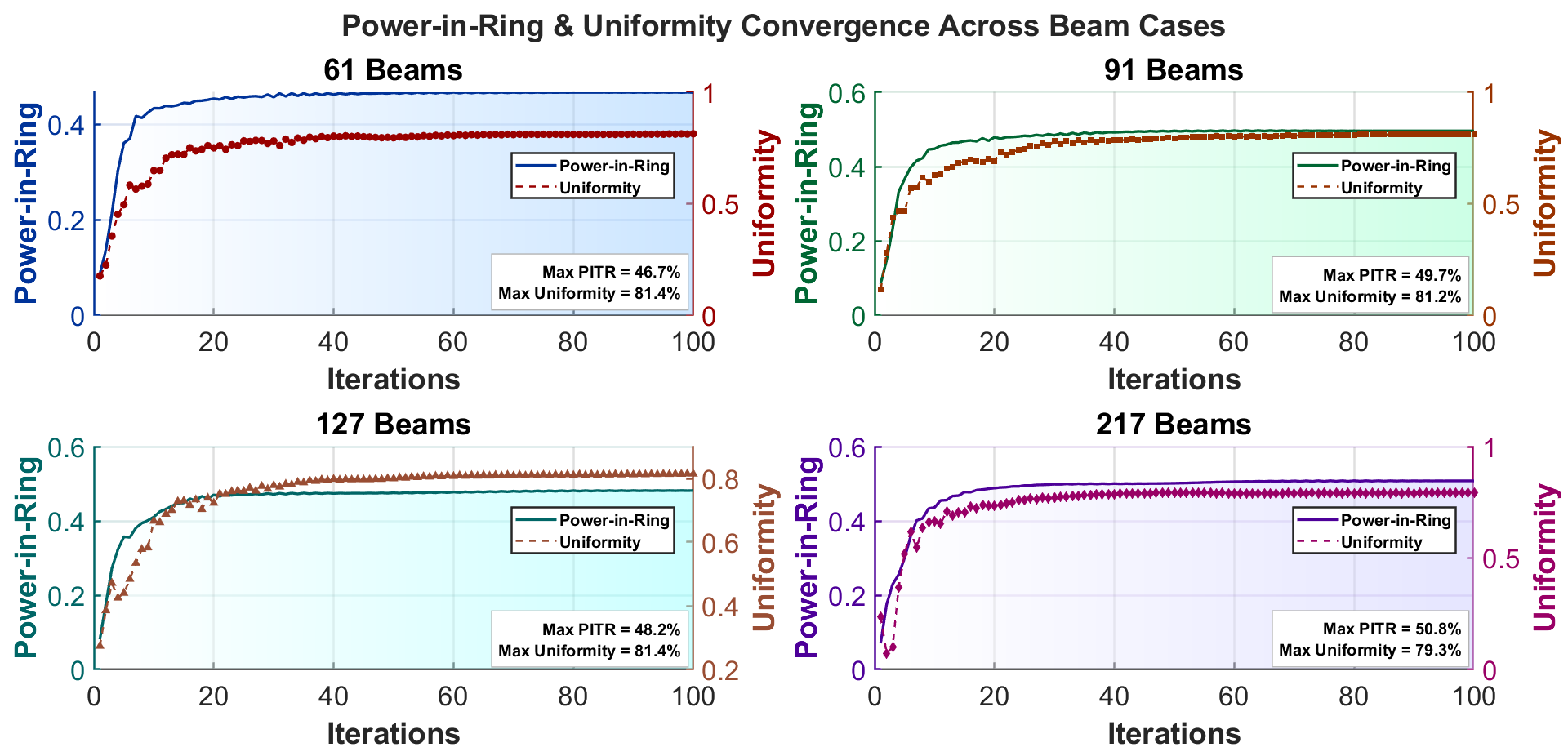}\label{fig:Ring_Combined_graphs_4_for_dirrent_beam_1}}

    \caption{\centering CBC convergence analysis using the Adagrad algorithm for ring shape: (a–d): Optimized far-field intensity profiles for 61, 91, 127, and 217 beams, (e): Convergence plots (iteration vs power-in-rectangle and uniformity) for different beams}
    \label{Ring_beam}
\end{figure*}
\begin{figure*}[p]
    \centering
     \subfigure[]{\includegraphics[width=0.24\linewidth]{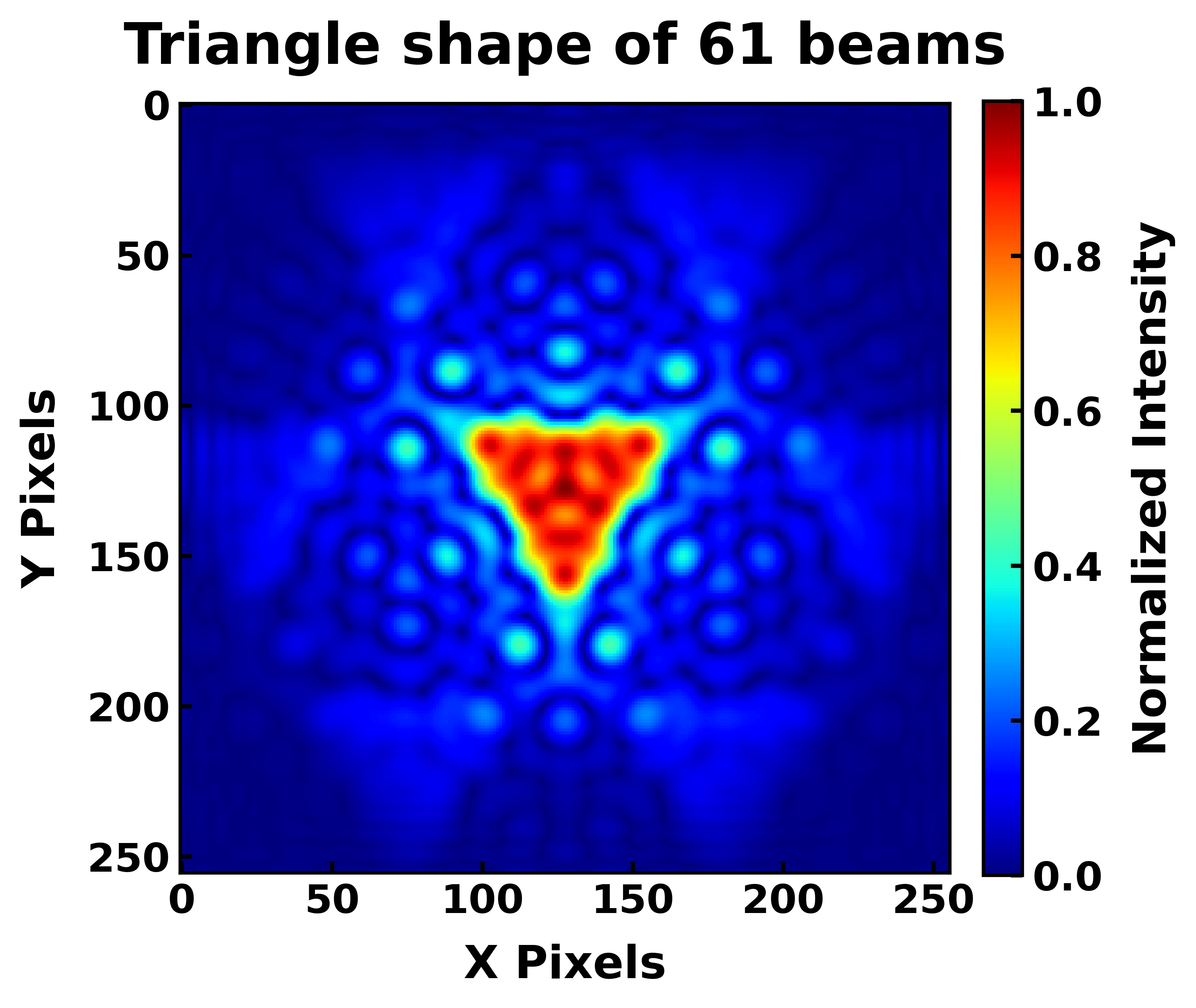}\label{fig:61_triangle_1}}
    \subfigure[]{\includegraphics[width=0.24\linewidth]{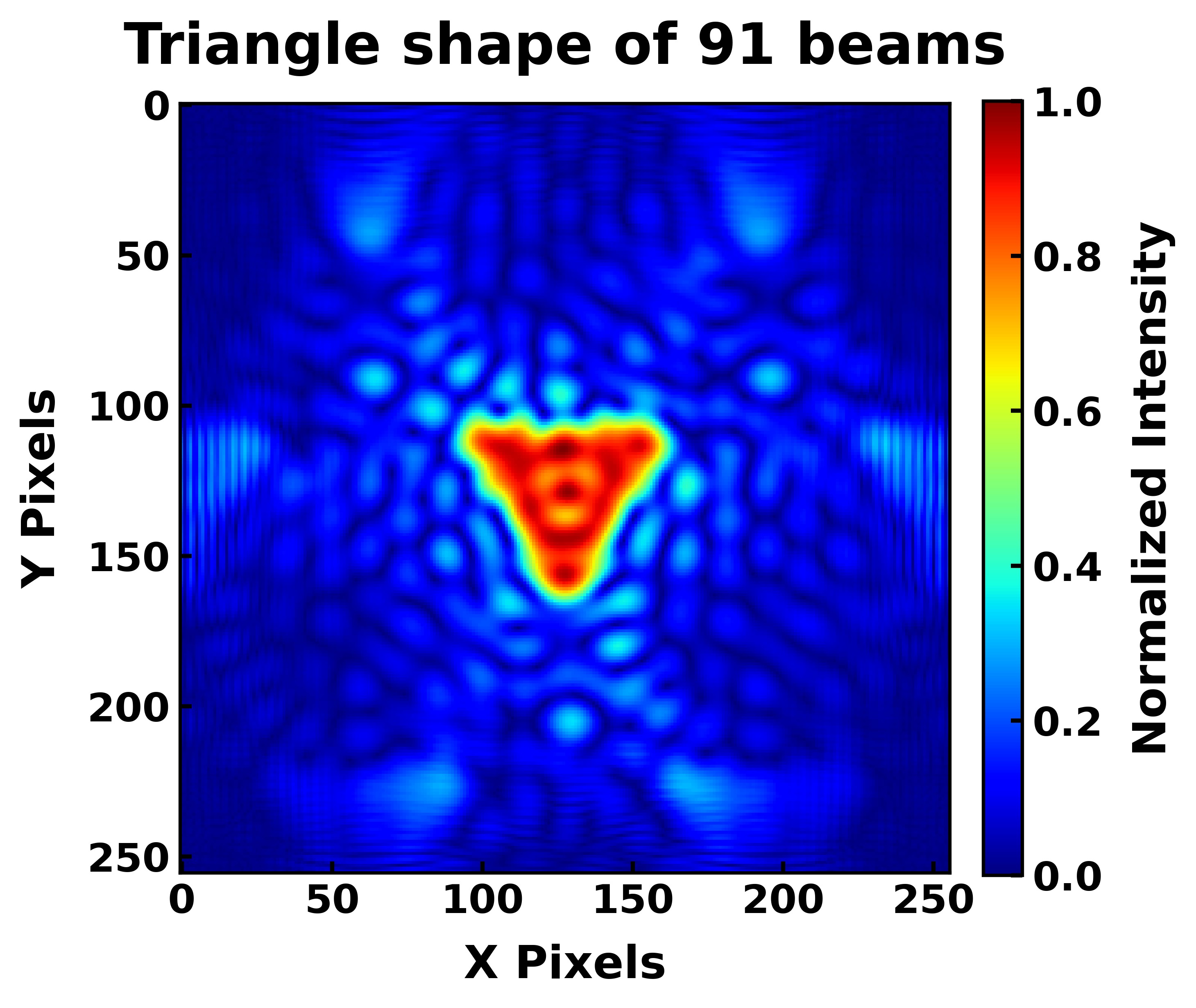}\label{fig:91_triangle_1}}
    \subfigure[]{\includegraphics[width=0.24\linewidth]{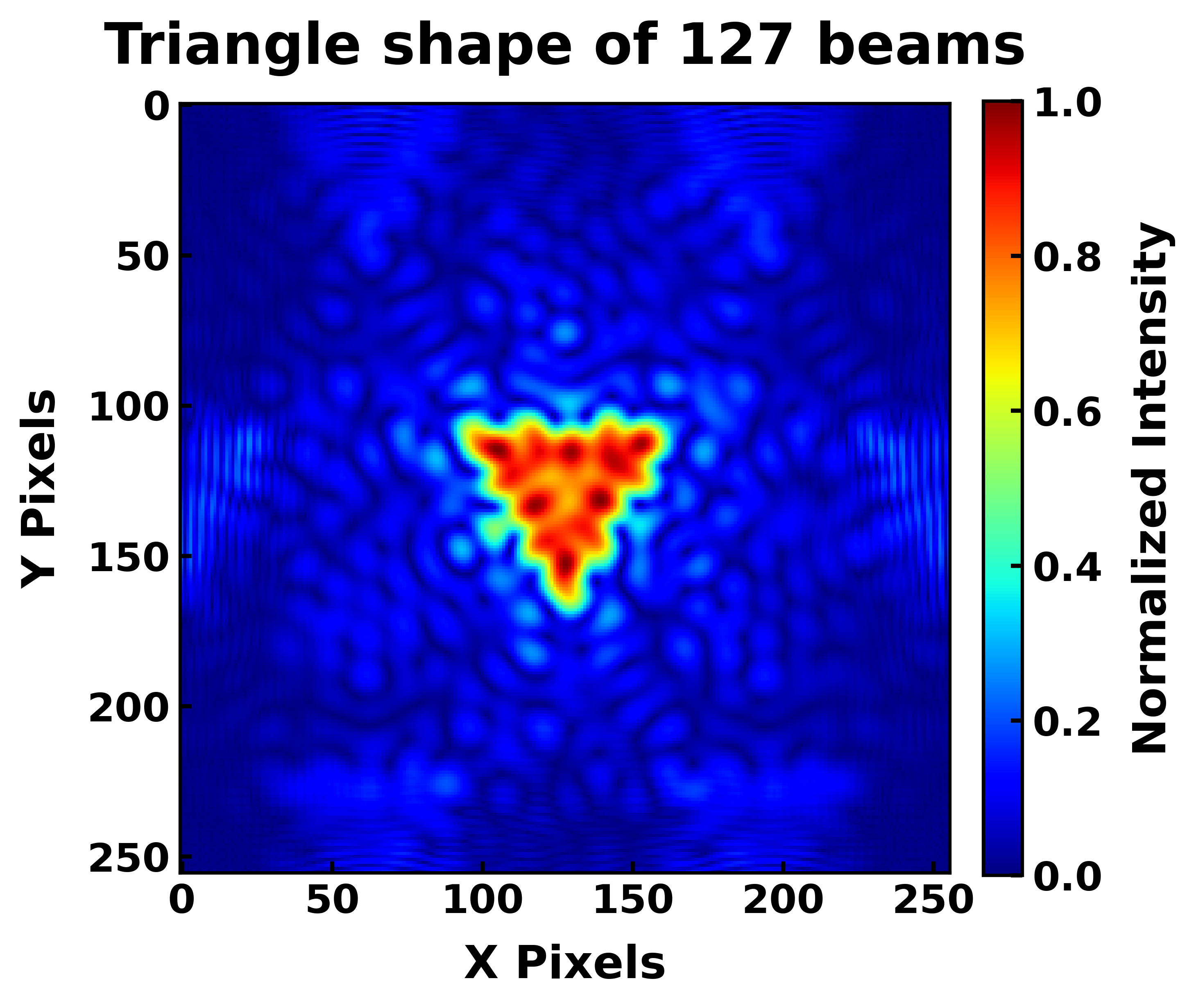}\label{fig:127_triangle_1}}
    \subfigure[]{\includegraphics[width=0.24\linewidth]{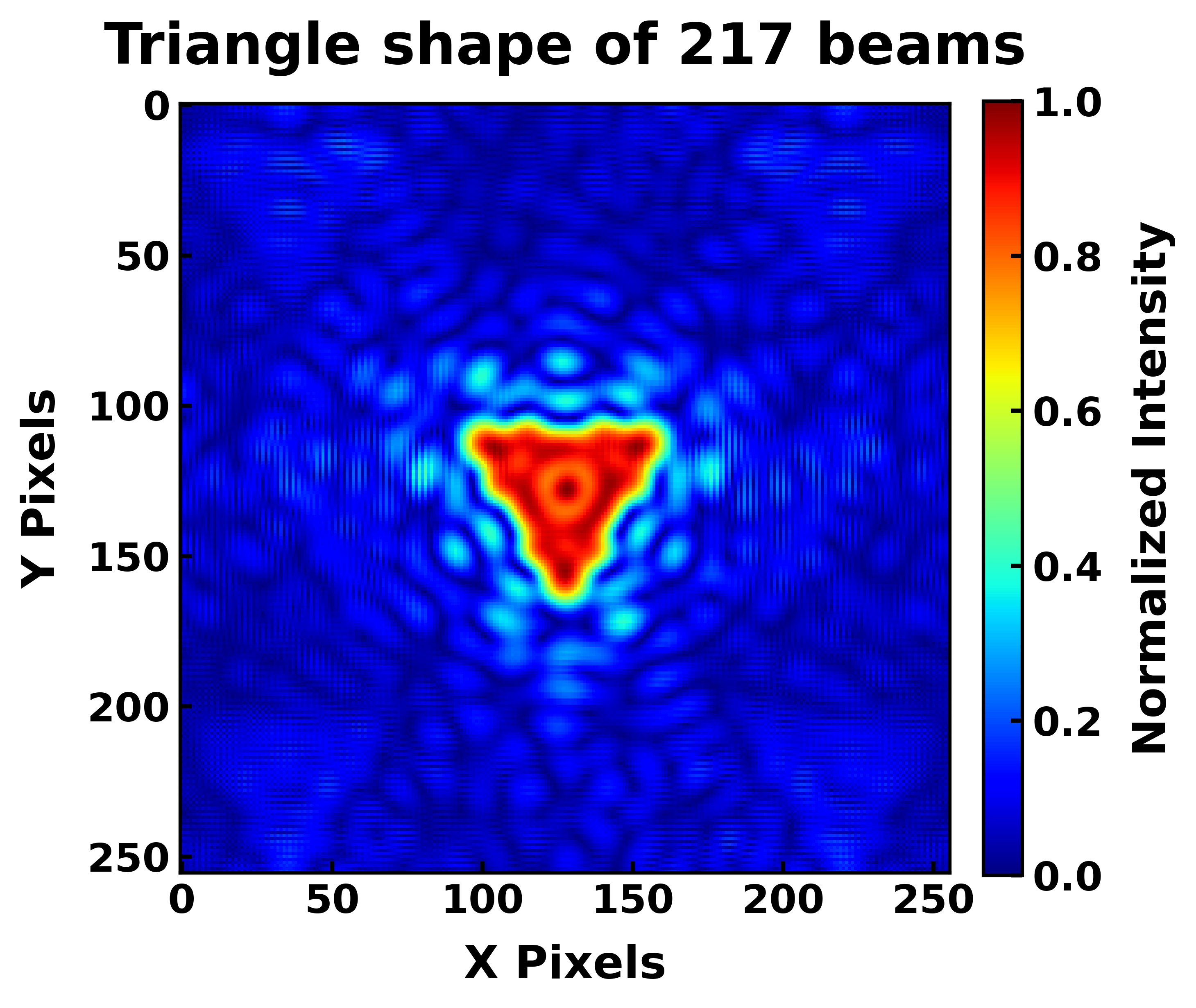}\label{fig:217_triangle_1}}
    \par\medskip 

    \subfigure[]{\includegraphics[width=0.95\linewidth]{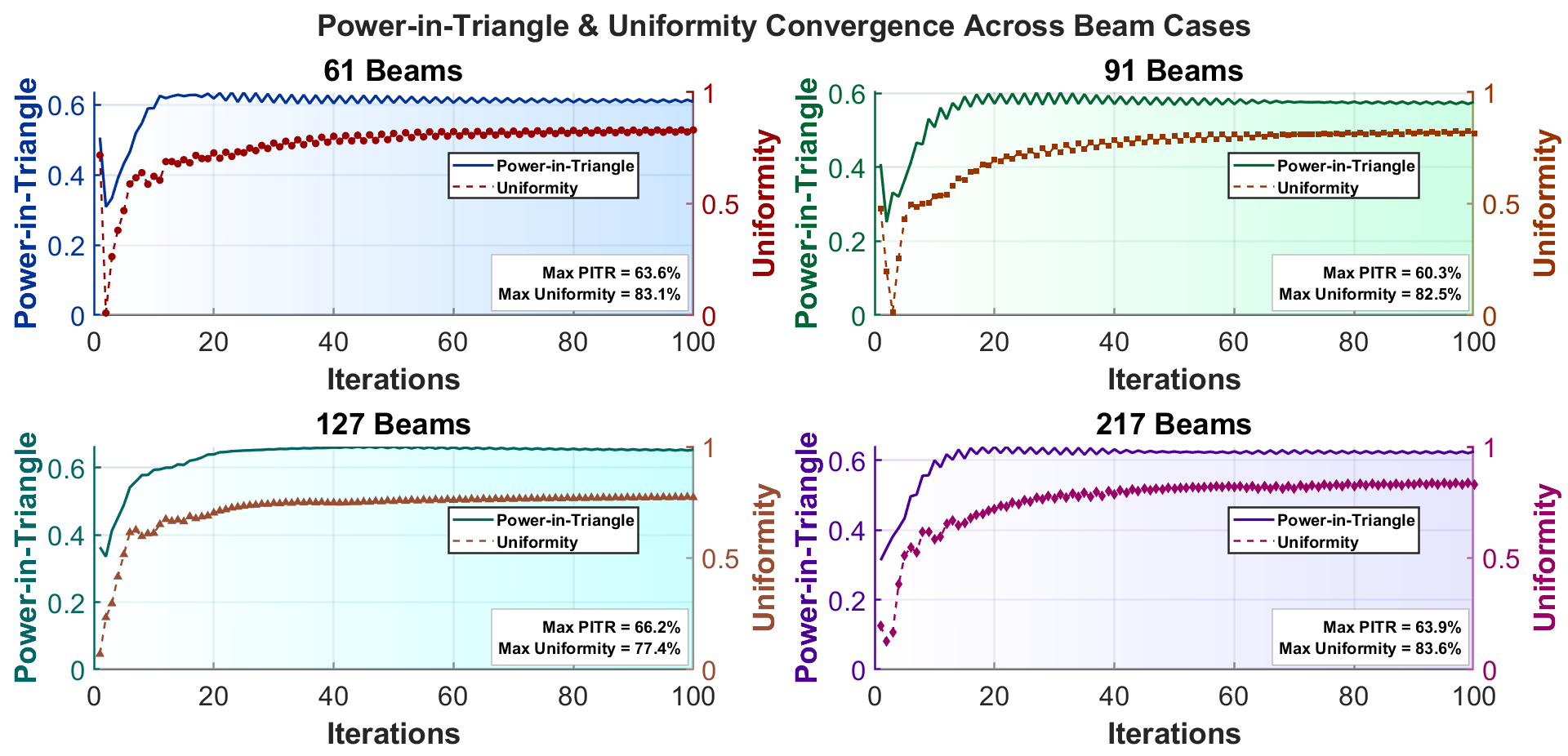}\label{fig:Triangle_4_combined_plot_different_beam_1}}

    \caption{\centering CBC convergence analysis using the Adagrad algorithm for triangle shape: (a–d): Optimized far-field intensity profiles for 61, 91, 127, and 217 beams, (e): Convergence plots (iteration vs power-in-rectangle and uniformity) for different beams}
    \label{Triangle_plot}
\end{figure*}
\begin{figure*}[p]

\centering
\fbox{\includegraphics[width=\linewidth]{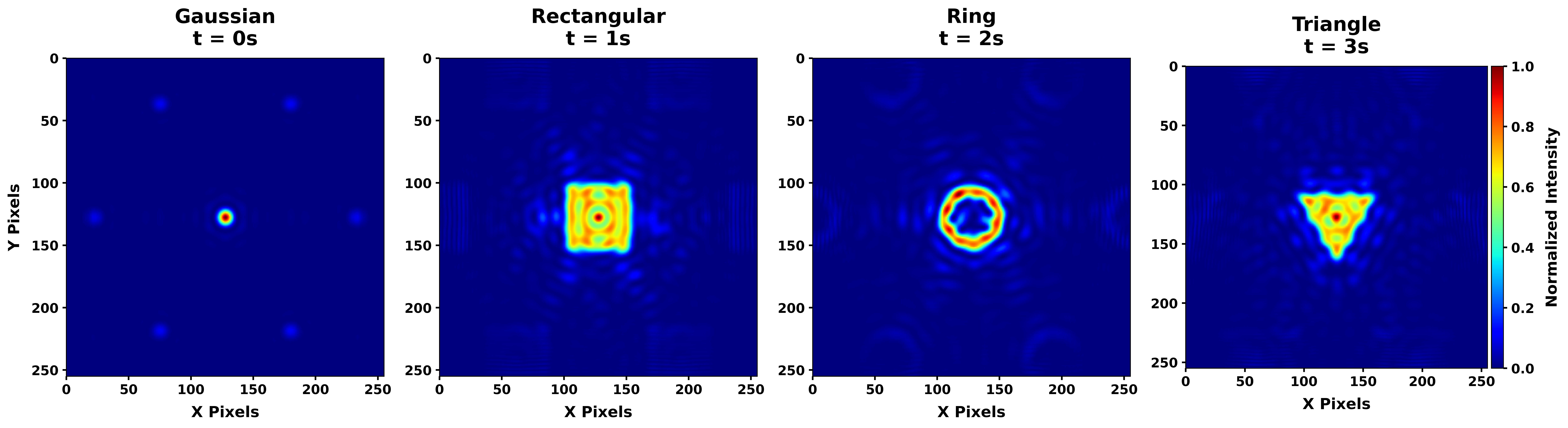}}
\caption{\centering Dynamic beam shaping at 1-second intervals using pre-locked phase masks, transitioning through Gaussian, rectangle, ring, and triangular patterns}
\label{fig:Dynamic_beam_shape_image}
\end{figure*}

\paragraph{Far-field model and per-pixel power}

Let the output electric field in the far field be $E_{\mathrm{out}}(x,y)$. The intensity distribution is

\begin{equation}
I(x,y) = |E_{\mathrm{out}}(x,y)|^2
\end{equation}
For a discrete grid with pixel size $\Delta x$, the per-pixel power is
\begin{equation}
P(x,y) = I(x,y)\Delta x^2
\end{equation}

\paragraph{Target mask}

A binary or soft mask $M(x,y)\in[0,1]$ specifies the desired support region of the far-field intensity. The optimization process aligns the beam pattern with this mask.

\paragraph{Power-in-target region (PITR)}

The power deposited inside the mask is
\begin{equation}
\mathrm{PITR}(M) = \sum_{x,y} P(x,y)M(x,y)
\end{equation}
Normalizing by the total power gives the fractional in-mask power,
\begin{equation}
\mathrm{PITRf}(M) =
\frac{\mathrm{PITR}(M)}{\sum_{x,y} P(x,y)}
\end{equation}
which is scale-invariant and independent of absolute beam intensity.

\paragraph{Uniformity}

To evaluate intensity evenness inside the mask, we define the normalized per-pixel power
\begin{equation}
p(x,y) = \frac{P(x,y)M(x,y)}{\mathrm{PITR}(M)}
\qquad
\sum_{x,y} p(x,y)=1
\end{equation}
Let $N_M=\sum_{x,y} M(x,y)$ denote the number of effective mask pixels. The masked mean and variance are
\begin{equation}
\mu_M=\frac{1}{N_M}\sum_{x,y} p(x,y)M(x,y)
\end{equation}
\begin{equation}
\sigma_M=\sqrt{\frac{1}{N_M}\sum_{x,y} M(x,y)(p(x,y)-\mu_M)^2}
\end{equation}
The uniformity metric is then
\begin{equation}
U_M = 1 - \frac{\sigma_M}{\mu_M+\varepsilon},
\qquad 
U_M \leftarrow \operatorname{clip}(U_M,0,1)
\end{equation}

where $\varepsilon \ll 1$ is a small regularizer, and a value of $U_M = 1$ represents perfect uniformity.

\paragraph{Center leakage (ring beams)}

For ring-type beams, any intensity at the center is unwanted. A secondary mask $M_{\mathrm{center}}(x,y)$ penalizes center leakage:
\begin{equation}
L_{\mathrm{center}} = 
\frac{\sum_{x,y} P(x,y)\,M_{\mathrm{center}}(x,y)}
{\sum_{x,y} P(x,y)}
\end{equation}

\paragraph{Composite merit function}

The optimization objective combines efficiency, uniformity, and leakage suppression:
\begin{equation}
J(M) = \mathrm{PITRf}(M)\cdot U_M \cdot \bigl(1 - L_{\mathrm{center}}\bigr)^{\gamma}
\end{equation}

where $\gamma\ge0$ controls the strength of center suppression. For rings, $\gamma>0$ enforces a dark core, while for rectangular and triangular beams $\gamma=0$.

\paragraph{Target mask definitions}

\begin{itemize}
  \item \textbf{Ring beam:} inner and outer radii $r_{\mathrm{R1}},r_{\mathrm{R2}}$ define
  \begin{equation}
M_{\mathrm{ring}}(x,y) =
\begin{cases}
1, & r_{\mathrm{R1}} \le \sqrt{x^{2}+y^{2}} \le r_{\mathrm{R2}}, \\[4pt]
0, & \text{otherwise}.
\end{cases}
\end{equation}

  Center suppression uses
  \begin{equation}
M_{\mathrm{center}}(x,y) =
\begin{cases}
1, & \sqrt{x^{2}+y^{2}} < 0.3\, r_{\mathrm{R1}}, \\[4pt]
0, & \text{otherwise}.
\end{cases}
\end{equation}

  \item \textbf{Rectangle beam:} with side length $L_{\mathrm{rc}}$,
  \begin{equation}
M_{\mathrm{square}}(x,y) =
\begin{cases}
1, & |x| \le \tfrac{1}{2}L_{\mathrm{rc}} \ \text{and}\ |y| \le \tfrac{1}{2}L_{\mathrm{rc}}, \\[4pt]
0, & \text{otherwise}.
\end{cases}
\end{equation}

  \item \textbf{Triangular beam:} equilateral triangle centered at origin with circumradius $R$,
  \begin{equation}
M_{\triangle}(x,y) =
\begin{cases}
1, & (x,y) \in \operatorname{conv}\{v_1, v_2, v_3\}, \\[4pt]
0, & \text{otherwise}.
\end{cases}
\end{equation}

  where $v_1,v_2,v_3$ are vertices at $0^\circ,120^\circ,240^\circ$.
\end{itemize}




We applied this framework to CBC arrays of 61, 91, 127, and 217 beams arranged hexagonally. Random initial phases were optimized to maximize $J(M)$ for three representative target profiles—rectangle, ring, and triangle as shown in figures~\ref{Square_plot}, \ref{Ring_beam}, \ref{Triangle_plot} respectively. Performance was evaluated using fractional power in the target region ($\mathrm{PITRf}$) and uniformity ($U_M$). Each target pattern has distinct relevance: rectangle beams provide uniform energy deposition for lithography or heating, ring beams enable central-void applications such as optical trapping or surgery, and triangular beams support structured illumination.  

The figures~\ref{Square_plot}, \ref{Ring_beam}, \ref{Triangle_plot} highlight two aspects of the results. Subfigures (a–d) in figures show the optimized far-field intensity distributions for different array sizes, confirming that CBC can reliably reproduce structured profiles even with hundreds of beams. Triangular beams form clean edges, rectangular beams exhibit sharp boundaries with consistent coverage, and ring beams maintain radial symmetry with a clear central void.

Subfigure (e) in figures~\ref{Square_plot}, \ref{Ring_beam}, \ref{Triangle_plot} presents the convergence behavior of fractional power in the target region and uniformity across different beam counts. In all cases, the optimization process steadily improves both metrics over iterations, eventually reaching stable values. While the specific convergence rate varies with array size, the final outcomes remain consistent, with each beam configuration achieving high power concentration and uniformity in the target shapes. Across all beam counts rectangular beams achieve up to 70\% power with uniformity above 83\% as shown in subfigure~\ref{fig:Ractangle_combined_4_different_beams_1}, ring beams stabilize near 50\% power with uniformity around 79–81\% as illustrated in subfigure~\ref{fig:Ring_Combined_graphs_4_for_dirrent_beam_1}, and triangular beams reach fractional power levels of around 62–63\% with uniformity near 83\% as evident from subfigure~\ref{fig:Triangle_4_combined_plot_different_beam_1}. These results confirm that while convergence speed depends on array size, the final shaped-beam quality remains high and robust.

Uniformity improves alongside power concentration, eventually stabilizing across all patterns and beam counts. This demonstrates that the optimized beams not only direct energy efficiently into the desired regions but also distribute it evenly—an essential property for practical applications requiring consistent illumination or controlled thermal deposition.

Overall, the results show that CBC arrays can reliably transform random initial phases into structured far-field intensity distributions with both high fidelity and balanced energy delivery. The method scales well with increasing beam numbers and preserves the geometric integrity of the target shapes, highlighting its potential for advanced applications in lithography, laser processing, optical trapping, and structured illumination.

\subsection{\textbf{Dynamic beam sequencing}}
In many advanced photonic workflows, especially multi-step processes such as additive manufacturing, different stages often require distinct beam patterns or energy distributions. Material deposition, for example, may benefit from a Gaussian beam to achieve controlled melt-pool formation, whereas surface finishing or precision post-processing may demand top-hat or ring-shaped profiles to ensure uniform energy deposition. Such dynamic reconfigurability is therefore essential for maintaining process efficiency, thermal stability, and spatial repeatability across varying operational requirements. By enabling rapid transitions between engineered far-field shapes, a dynamically reconfigurable beam-shaping system improves overall process consistency and allows real-time adaptation to changing task demands.

To demonstrate this capability, we developed a phase lookup table containing pre-computed phase values for various target beam profiles, such as ring, rectangle, and triangular patterns. These phase values were first locked by running the optimization independently for each desired shape. During the dynamic demonstration, these stored phase patterns were sequentially applied to the beam combining system at fixed time intervals to switch the far-field pattern. This approach emulates a dynamic system and highlights how precomputed phases can be effectively used for rapid beam reconfiguration. The accompanying image~\ref{fig:Dynamic_beam_shape_image} shows the far-field intensity profiles at different time steps, each representing a distinct beam shape, 


The transitions are smooth and closely match the intended patterns, confirming that a pre-computed phase map is effective for applications where real-time optimization would be computationally expensive.

What makes this approach particularly valuable is that it relies solely on computational control; no physical beam shaper or deformable mirrors are needed in this method. 
Furthermore, because the optimization process generalizes well to different target patterns, our method offers a flexible framework for on-demand beam shaping. Instead of designing custom optics for each desired profile, one could simply redefine the target shape and let the optimizer re-tune the phases accordingly.

In terms of practical implications, these findings open new directions for beam control in laser systems. Whether it is maximizing efficiency in directed energy platforms or achieving fine precision in laser manufacturing. The ability to dynamically shape beams to high accuracy is a game changer. Additionally, the results demonstrate that even with an increased number of beams, which traditionally adds complexity, the system can still maintain excellent performance, both in terms of power delivery and beam quality.

\section{\textbf{Summary and conclusion}}

The rapid development of high-power laser systems has created a strong demand for beam-engineering platforms that are both reconfigurable and scalable. Modern applications such as additive manufacturing, precision materials processing, optical manipulation, and directed-energy systems require optical fields that can be reshaped, steered, and adapted in real time. Meeting these requirements calls for methods that go beyond conventional static optics and provide intelligent, programmable control over high-power beams.

In this work, we have developed an adaptive coherent beam combining (CBC) framework that addresses these emerging needs by enabling both precise beam steering and flexible beam shaping through computational phase control. The core of our approach lies in formulating CBC as an optimization-driven light-engineering problem, where the phases of many laser channels are adaptively tuned to synthesize user-defined far-field intensity distributions. To overcome the challenges associated with large beam arrays, we employed the Adagrad optimizer, whose per-parameter learning adaptation significantly improves convergence stability and scalability compared to fixed-rate methods.
Three complementary beam engineering strategies were demonstrated:

\begin{enumerate}
\item \textbf{Beam steering and beam shaping through sequential steering:} The work first establishes precise beam steering by directing the CBC focus to selected target points across the far-field plane. Building on this capability, complex geometries were generated by sequentially steering the beam across a set of predefined positions that together outline the desired pattern. At each point, the phase was optimized to produce a clean and well-focused spot, and these optimized states were combined in sequence to form the complete shape. This sequential steering approach offers an effective route for producing intricate patterns by assembling multiple localized focal positions.

To ensure uniform brightness within the constructed geometry, a dwell-time correction was introduced: such that,  the pixels  receiving naturally higher intensity were assigned shorter exposure durations, while weaker points were held longer. This adaptive weighting resulted in a uniformly illuminated and well-balanced far-field pattern.

\item \textbf{Structured static beam shaping through adaptive phase optimization:} Predefined beam profiles were used as targets for the optimization, allowing the CBC system to directly form well-defined far-field shapes such as rings, rectangles, and triangles. The approach delivered strong power concentration and high uniformity across all tested patterns, demonstrating its capability to generate stable, structured beam patterns with balanced energy distribution. 

\item \textbf{High-speed dynamic beam sequencing through pre-computed phase switching:} To support applications where beam profiles must change rapidly, we generated a set of pre-optimized phase states for multiple target patterns and switched among them in real time. This approach enables smooth and accurate transitions between shapes such as Gaussian, rectangle, ring, and triangle without relying on mechanical motion or additional optical components. Because the optimization is performed offline, the system can deliver fast, on-demand beam reconfiguration while maintaining high beam quality and power concentration. These results highlight the potential of CBC-based phase switching as a practical and scalable solution for next-generation laser systems requiring rapid, precise, and energy-efficient beam adaptation.  
\end{enumerate}

Together, these limitless capabilities of CBC establish a comprehensive and versatile framework for on-demand beam engineering. The ability to electronically sculpt high-power beams without mechanical scanning or custom optics makes the approach especially valuable for laser-based manufacturing, precision processing, optical manipulation, and other fields where flexible, reconfigurable light fields are essential.

Overall, the results presented in this work establish a clear foundation for advancing CBC toward more capable, application-driven laser systems. Rather than serving as a single demonstration, the framework illustrates how phase-optimized beam control can be structured, extended, and adapted to a broad range of operational goals. Its modular nature allows users whether in research laboratories or industrial environment to adopt and customize the elements most relevant to their needs, from precise steering to structured shaping or rapid profile transitions. Although the present study focuses on computational demonstrations, our ongoing efforts aim to experimentally validate the capabilities of the CBC system for broad manufacturing applications, further reinforcing its practical relevance and readiness for real-world deployment. Aware of the rapid developments in CBC across industrial, space, and defense applications, our results are expected to support and guide the development of future adaptive and programmable laser systems.


\section*{\textbf{Acknowledgment}}

The authors acknowledge Extramural Research \& Intellectual Property Rights (ER\&IPR/D(R\&D)/1865), CEFIPRA/IFCPAR~(IFC/7148/2023), and UKIERI-SPARC~(3673) for financial support through research projects. Additionally, KN thanks the Anusandhan National Research Foundation, India, for support through the Core Research Grant (CRG/2023/008068).

\ifCLASSOPTIONcaptionsoff
  \newpage
\fi



%


\bibliography{reference.bib}
%








\end{document}